\documentclass[prr,preprint,floatfix,epsf,psfig]{revtex4}
\pdfoutput=1 %% this is very important command to upload pdf figures in condmat
\usepackage{epsf}
\usepackage{graphicx}
\renewcommand{\figurename}{FIG.}
\usepackage[hypertexnames=false,linktocpage=true]{hyperref}
\hypersetup{colorlinks=true,linkcolor=blue,anchorcolor=blue,citecolor=blue,menucolor=blue,filecolor=blue,urlcolor=blue}

\begin{document}
%	\title{Thick layer of Sn on $i$-Al-Pd-Mn: a quasicrystal clathrate?}
%	\title{Thick quasicrystalline layer of Sn on $i$-Al-Pd-Mn: a novel clathrate structure?}
\title {Quasiperiodic ordering in thick Sn layer on $i$-Al-Pd-Mn: A possible quasicrystalline clathrate}
%\title{Thick layer of Sn on i-Al-Pd-Mn: quasicrystal clathrate? \\OR\\ Quasiperiodic ordering in thick Sn layer on $i$-Al-Pd-Mn: a quasicrystal clathrate?}
%\title{A new clathrate based quasicrystal model suggested for a thick layer of Sn on $i$-Al-Pd-Mn}
%\title{A new clathrate based quasicrystal observed in thick layer of Sn on $i$-Al-Pd-Mn}
\author{Vipin Kumar Singh$^{1\dagger}$, Marek Mihalkovic$^{2*\dagger}$,
	Marian Kraj\v{c}\'i$^{2**\dagger}$, Shuvam Sarkar$^{1\dagger}$, Pampa Sadhukhan$^{1\dagger}$, M. Maniraj$^{1\dagger}$, Abhishek~Rai$^{1\dagger}$,    Katariina Pussi$^{3}$,  Deborah L. Schlagel$^{4}$,  Thomas  A. Lograsso$^{4}$,   Ajay Kumar Shukla$^{5}$,  Sudipta Roy Barman$^{1\ddagger}$}

\affiliation{$^{1}$UGC-DAE Consortium for Scientific Research, Khandwa Road, Indore - 452001, Madhya Pradesh, India}
\affiliation{$^{2}$Institute of Physics, Slovak Academy of Sciences, D$\acute{u}$bravsk$\acute{a}$ cesta 9, SK-84511 Bratislava, Slovak Republic}
\affiliation{$^{3}$ LUT School of Engineering Science, P.O. Box 20, FIN-53851 Lappeenranta, Finland}
\affiliation{$^{4}$Division of Materials Sciences and Engineering, Ames Laboratory, Ames Iowa  500011-3020, USA}
\affiliation{$^{5}$CSIR-National Physical Laboratory, Dr. K. S. Krishnan Road, New Delhi - 110012, India}
\affiliation{$^{\dagger}$These authors have  contributed equally to this work.}
%\affiliation{$^{*,**,\#}$ Corresponding authors}
	
\begin{abstract}
    Realization of an elemental  solid-state quasicrystal has remained a distant dream so far in spite of extensive work in this direction for almost two decades.  In the present work,  we report the discovery of   quasiperiodic ordering in a thick layer of elemental Sn grown on icosahedral ($i$)-Al-Pd-Mn.   The  scanning tunneling microscopy (STM) images and the low energy electron diffraction patterns of the  Sn layer show specific structural signatures that portray quasiperiodicity but are distinct from the substrate.  Photoemission spectroscopy reveals the existence of the pseudogap around the Fermi energy  %ing in $i$-Al-Pd-Mn further deepens for the first Sn monolayer, and persists 
   ~up to the maximal Sn thickness. 
  The structure of the Sn layer  is modeled as
   a novel form of quasicrystalline clathrate
   %   {\bf We \underline {propose a quasicrystal model} for the Sn layer to be a novel form of quasiperiodic clathrate 
   ~on the basis of %by bringing forward 
   	~multiple supporting evidences: Firstly, from {\it ab-initio} total energy evaluation, the  energy of bulk Sn clathrate quasicrystal is lower than   the high temperature metallic $\beta$-Sn phase, but higher than the low temperature $\alpha$-Sn phase. A comparative study of the free slab energetics shows that surface energy favors clathrate over $\alpha$-Sn up to about 4 nm layer thickness, and matches $\beta$-Sn for narrow window of slab thickness  of 2-3 nm. %SRB: changed this sentence to make it more clear, if you do not like can revert back.
      ~ Secondly, the bulk  clathrate exhibits gap opening near Fermi energy, while the free slab form exhibits a pronouced pseudogap, which explains the pseudogap observed in photoemission. Thirdly, the STM images exhibit good agreement with clathrate model.
  %       are well reproduced  by the clathrate model. 
  ~Finally, we establish the adlayer-substrate  compatibility based on very similar (within 1\%) the cage-cage separation in the Sn clathrate and the pseudo-Mackay cluster-cluster separation  on the $i$-Al-Pd-Mn surface. %(1.28 nm) enjoy magic consistency 
   Furthermore, the nucleation centers of the Sn adlayer on the substrate are identified and these are shown to be  a valid part of the Sn clathrate structure. Thus, based on  both experimental and {\it ab-initio} density functional theory  calculations, we  propose %establish 
   ~that 4~nm thick Sn adlayer deposited on 5-fold surface of $i$-Al-Pd-Mn substrate is in fact a metastable realization of elemental, clathrate family quasicrystal.  %Sn thus exhibits non-pseudomorphic growth to a thickness where the substrate potential is negligible.  The role of the substrate is solely to suppress the initial formation of any stable crystalline forms of Sn.}%SRB: I think it is 2-3 nm, rather than 3 nm, in fact at 2 nm I see that clathrate is just equal to beta Sn**

\end{abstract}
%\pacs{~71.23.Ft, %%Quasicrystals, part of 71.00.00 Electronic structure of bulk materials 
%	79.60.Dp, %photoemission :	from adsorbed layers, 
%	61.05.jh, %Low energy electron diffraction (LEED), 
%	68.37.Ef} %Scanning tunneling microscopy (STM) :	in study of surface structure
\maketitle

%\newpage
%{\begin{center} {\bf Significance statement} 
%\end{center}}
%	Although quasicrystals have been discovered in 1984, until date, the quasicrystalline phase could not be identified within single-elemental systems, and even metastable elemental quasicrystal phase did not appear possible. In the present work, we bring multiple evidences -- both experimental and based on $ab-initio$ calculations -- that 4~nm thick Sn adlayer grown on $i$-Al-Pd-Mn quasicrystal substrate is in fact metastable realization of elemental, clathrate family quasicrystal. From $ab-initio$ total energy calculations, unsupported 3-4 nm thick slabs with pentagonal symmetry and clathrate structure have lower energy than crystalline $\alpha$-Sn, suggesting that formation of quasiperiodic clathrate of large thickness, where the  of the substrate is solely to suppress initial formation of crystalline Sn. 
%\newpage 

\section{Introduction}
The mathematical concept of aperiodic ordering  and the discovery of quasicrystals  brought paradigm shift in crystallography\cite{Shechtman_prl84,Penrose74}. Quasiperiodicity has been  observed in naturally occurring minerals\cite{Bindi09}, different inter metallic compounds\cite{Tsai00}, binary nano-particle super lattices\cite{Talapin09} colloidal quasicrystal\cite{Fischer11}, oxide thin film\cite{Forster13} and large molecular assemblies\cite{Xiao12,Ye16}.  A recent exciting finding is the nontrivial topological property of quasiperiodic systems\cite{Kraus16,Zhou19}.% originating from higher dimensions\cite{Kraus16,Miguel16,Zhou19}. 
~The technological importance of quasicrystals arise from their low frictional coefficient, heat insulation and  photonic band gap\cite{Park05,Dubois12}.  

However, the chemical complexity of  quasicrystals discovered so far hinders their applicability and  basic understanding of their unusual properties.  An elemental quasicrystal  would be best suited for this purpose, but a single-element bulk quasicrystal has not been yet observed.  Efforts in this direction for over a decade have involved the use of  quasicrystalline substrates as a template to grow  quasiperiodic elemental  layers\cite{Franke_prl02, Shimoda_jns04, Sharma_prb05, Krajci_Snprb05,Shukla_prb06, Ledieu_prb08,Smerdon_prb08,LedieuPbAPMprb09,Sharma_prb08,Shukla_prb09,Krajci_PbAPM_prb10, Sharma13, Hars_ss18}. Pseudomorphic growth of one monolayer (ML) quasiperiodic Sn has been reported  in the past on  decagonal ($d$)-Al-Ni-Co\cite{Shimoda_jns04} and icosahedral ($i$)-Al-Cu-Fe\cite{Sharma_prb05}. A density functional theory (DFT)  study predicted pseudomorphic  growth of quasiperiodic Sn on $i$-Al-Pd-Mn\cite{Krajci_Snprb05}. In most of the  studies in the past,  only the first wetting layer has been reported to exhibit five-fold quasiperiodicity with a structure similar to the substrate.  Pb/$i$-Ag-In-Yb showed five-fold growth isostructural with the substrate and pentagonal motifs with maximum height of about 0.7 nm were identified\cite{Sharma13}.
 It may be noted that although the thickness was substantial, five-fold quasiperiodicity was not observed in 25 ML Cu/$i$-Al-Pd-Mn\cite{Ledieu_prl04}  or in 1.8~nm thick Ag on GaAs\cite{Smith}, rather aperiodic modulation of row separation were observed  only in one direction.

It is interesting to note that elements from group XIV of the periodic table have
specific propensity to form pentagonal or even icosahedral structures.
 Although Sn exhibits a  diamond type structure ($\alpha$-Sn, space group Fd$\overline{3}$m) at low temperature and a tetragonal structure ($\beta$-Sn, space group I4$_1$/amd) above 286~K,  it is well known that Sn  can form $sp^3$-bonded
\textit{clathrate} structures whose canonical constituents are four
kinds of cages that are bounded exclusively by pentagonal and
hexagonal faces. Out of the four canonical cages, the most abundant
one in common clathrate crystals is the dodecahedral cage with
icosahedral symmetry, bounded exclusively by 20 pentagonal
faces, in contrast to purely hexagonal diamond structure (see Fig.\ref{cages} in Section III.D).
Intriguingly, in their computer simulation Engel \textit{et	al.}\cite{qcla} observed spontaneous formation of three icosahedral phases, and the ``loosely packed'' phase can be regarded as an
imperfect realization of the icosahedral quasicrystal clathrate. Last but not the least, the dodecahedron cage found in clathrates is identical with the second shell of the so called mini-Bergman clusters, playing fundamental role in $i$-Al-Pd-Mn structure.

In this work, we observed that Sn deposited on $i$-Al-Pd-Mn surface retains quasiperiodic ordering up to maximal achievable coverage leading to nearly 4 nm thickness, indicating the existence of the first intrinsic - although apparently metastable - monoatomic realization of a quasicrystal. We model it as a quasicrystal structure
belonging to the clathrate family strictly maintaining tetrahedral coordination consistent with $sp^3$ bonding.  We justify plausibility of the model by energetic considerations within the DFT framework, based on the structural comparison with experimentally observed motifs, and from the character of electronic density of states (DOS) near Fermi energy.

\section {\bf Methods}
%\subsection{Substrate surface and Sn layer preparation}
The single grain fivefold $i$-Al-Pd-Mn quasicrystal surface was prepared by repeated cycles of Ar$^+$ ions sputtering and annealing to about 900-930 K for 2$-$2.5 h. The  cycles were repeated until sharp fivefold low energy electron diffraction (LEED) pattern was observed.  Sn (99.99\% purity) was evaporated from a water cooled Knudsen evaporation cell\cite{Shukla_rsi04}. The evaporation cell was operated in the temperature range of 1073$-$1163 K at a  pressure of  2$\times$10$^{-10}$ mbar during the deposition. For the thick layer growth, we used  an effective deposition rate of  0.06 to 0.4~ML/min  in  repeated cycles of 1 min deposition  followed by 1 min waiting time. Sn deposition was carried out with the substrate close to room temperature (RT$\approx$ 300- 330~K) and  at 150 K. The temperature of the substrate was  measured using a K-type thermocouple.

%	\subsection{Experimental methods}
The  STM  measurements were carried out at a base pressure of 2$\times$10$^{-11}$ mbar using a variable temperature STM   from  Omicron GmbH.% STM   Resolution: 0.1 A (vertical), 1 A (lateral)
~STM measurements were performed at room temperature (RT) and 80 K in the constant current mode using a tungsten tip that was cleaned by field 	emission, sputtering and voltage pulse method.  The tip was biased and the sample was kept at the ground potential. It may be noted that same kind of tiles and motifs  are observed for different bias voltages ranging from  -2.6 to 2.8 V for different coverages, which demonstrate that the features are topography related.  The zero in the vertical scale of a STM image is set at the bearing height, which is  the most dominant height value, based on a height distribution histogram.  
%~The STM image analysis has been performed using the SPIP software from Image Metrology. 
~LEED was performed using  4-grid rear view  optics from OCI Vacuum Microengineering.  

The ultraviolet photoemission spectroscopy (UPS) measurements at 150 K were performed using R4000 electron energy analyzer with wide angle lens% from Scientaomicron%at RT using Phoibos  analyzer  from Specs GmbH   
~using  21.2  eV photon energy incident at 45$^{\circ}$ with respect to ($w.r.t$) the surface normal and measured in normal emission, with an analyzer acceptance angle of $\pm$15$^{\circ}$ in transmission mode and a resolution of 50 meV. 	The Fermi edge was measured on a metallic sample in electrical contact with the specimen. X-ray photoelectron spectroscopy (XPS) measurements were performed with 1253.6 eV (MgK$\alpha$) photon energy with a  resolution of 0.8 eV using Phoibos 100 electron energy analyzer. The thickness of the Sn adlayer has been determined  by STM and also by XPS from the intensities of Sn 3$d$ and Al~2$p$ core level peaks using the well known relations  $I_{Sn3d}$= $I_{Sn3d\infty}$(1- exp(-$d$/$\lambda$\,cos$\theta$)) and $I_{Al2p}$= $I_{Al2p\infty}$exp(-$d$/$\lambda$\,cos$\theta$) where  $d$ is the thickness, $\theta$ is the emission angle= 55$^{\circ}$ and  $\lambda$ is the mean free path. $\lambda$ for Sn 3$d$ (14.2\AA~) and Al 2$p$ (19.6\AA) are consistent with Ref.~\onlinecite{TPP2M}.

%\subsection{Density functional theory}
The DFT calculations have been carried out using the Vienna $ab-initio$ Simulation Package	(VASP)\cite{VASP1} by performing an iterative solution of the Kohn--Sham equations of DFT within a plane wave basis.  We use projector augmented wave potentials\cite{PAW} in the PW91 generalized gradient approximation\cite{PW91}. For both bulk and free slab structures, cohesive energies are converged to less than 1 meV/atom with basis set containing plane waves with a kinetic energy up to $E_{\rm cut-off}$= 150 eV.  The self-consistency iteration were stopped when total energies are converged to within 10$^{-6}$ eV. All the structures, bulk models as well as slabs, were fully relaxed without constraints, and including the cell parameters. $k$-point meshes have been converged to  satisfy for each mesh dimension $\alpha$~$N_{k\alpha}\times c_\alpha \sim$100, where $c_\alpha$ are edges of the periodic cell.

\begin{figure*}[t] 
	\includegraphics[width=210mm,keepaspectratio]{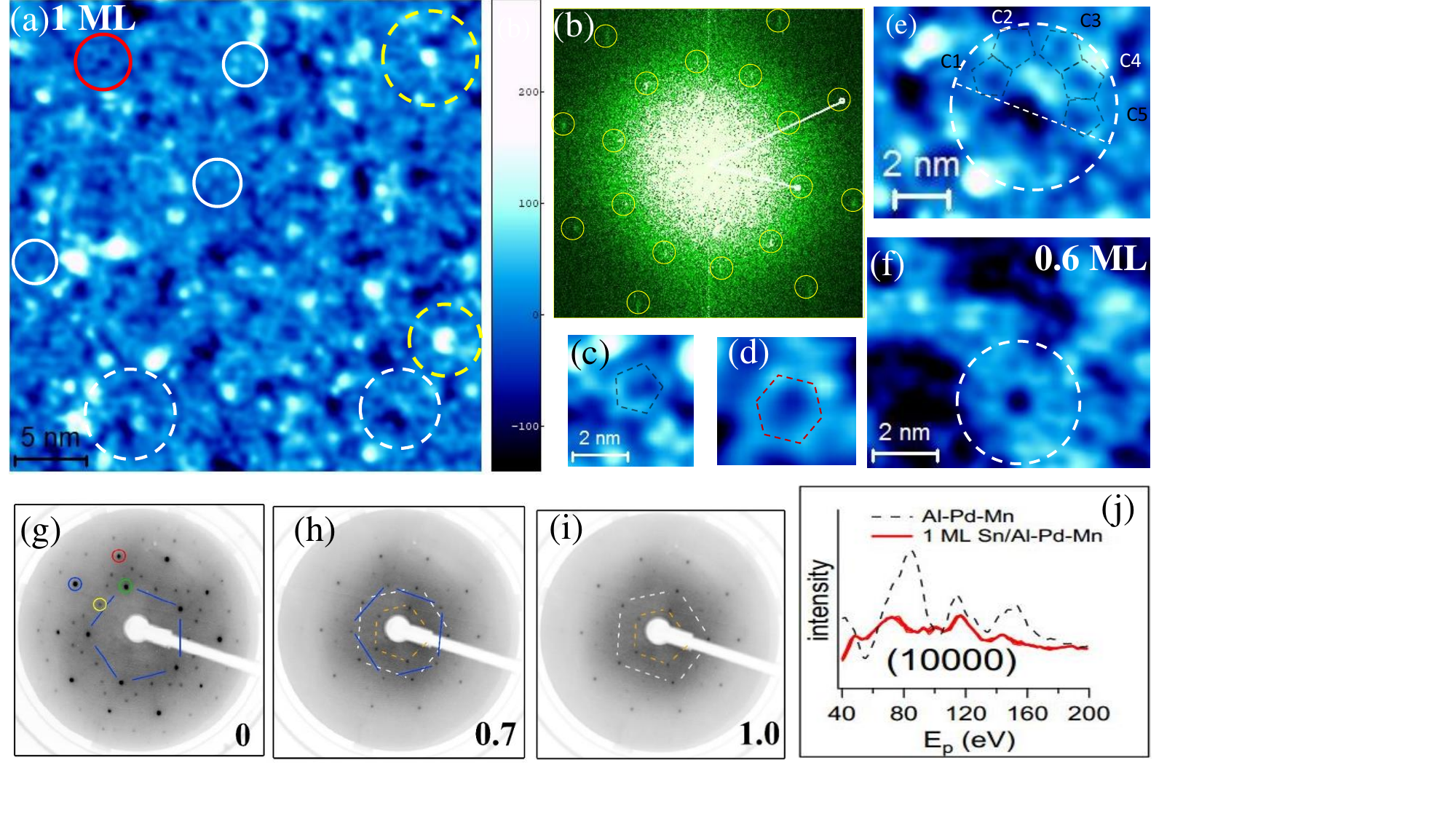}
	% Source: analysis/STM/Figures/STM_PPT_b02_12_2016_ver8_ps.ppt;Slide17 saved as pdf
	%1_Fig2_Slide17_ver
	%Source of (A): STM/Figures/PS_STM_images_for_Paper/IBM[17-1]25-Sep_1ML_1_5V_0_2nA_InvFFT_zoom36_wb.jpg from SPIP loaded in /analysisSTM/Figures/STM_PPT_02_12_2016_ver7 _ps.ppt, inserted circles, etc, grouped them, put in Slide16.
		
	\vskip -5mm
	\caption{ %{\bf Scanning tunneling microscopy up to 1st Sn layer}: 
		(a)  High resolution STM  topography image with tunneling current ($I_T$= 0.2~nA) and bias voltage ($U_T$= 1.5~V) for 1 ML Sn/$i$-Al-Pd-Mn  at room temperature (RT);  various motifs are marked  by circles. The color scale with the height  in picometers is shown on the right, zero corresponds to the bearing height. (b) Fourier transform (FT)  of  the STM image of 1 ML Sn/$i$-Al-Pd-Mn, the spots are highlighted by yellow circles.  The (c) Pentagonal ($P$) and (d) Hexagonal tiles are shown in expanded scale. (e) The crown motif enclosed by white dashed circle  formed by $P$-tiles  marked $C1$ through $C5$ (dashed black lines). (f)  0.6 ML Sn/$i$-Al-Pd-Mn with $I_T$= 0.6~nA, $U_T$= 2.8~V. The STM images are shown  after low-pass FT filtering.  LEED patterns  at beam energy of 81 eV for (g) $i$-Al-Pd-Mn (h) 0.7 ML Sn/$i$-Al-Pd-Mn and (i) 1 ML Sn/$i$-Al-Pd-Mn in  inverted scale, coverage is  indicated in ML at the bottom right corner.   (10000), (000$\bar{\it 1}$0), (00$\overline{\it 1}$\,$\overline{\it 1}$0) and (11000) LEED spots are highlighted by green, yellow, red and blue circles, respectively in (g). (j) Experimental IV curve for (10000) spot of 1  ML Sn/$i$-Al-Pd-Mn compared to that of $i$-Al-Pd-Mn.} 
		\label{stm}
\end{figure*}

\section{RESULTS AND DISCUSSION}
\subsection{STM and LEED of monolayer Sn on $i$-Al-Pd-Mn}
We begin our discussion by a high resolution STM image of  1 ML  Sn/$i$-Al-Pd-Mn in Fig.~\ref{stm}(a). It is clearly quasiperiodic, as demonstrated by its Fourier transform (FT) that  exhibits  two sets of spots of tenfold symmetry highlighted by yellow circles [Fig.~\ref{stm}(b)], whose distance from the center (white lines) is in the ratio of $\tau$, where  $\tau$ is the golden mean ($\approx$\,1.618).  The most prevalent tile in the Sn layer is a pentagon  with  a central dark region [Fig.~\ref{stm}(c)], as shown by black dashed lines. The  length of the sides of these pentagonal ($P$)-tiles is around 0.7~nm, as shown by its length distribution in Fig.~S1(a) of the Supplementary Material (SM)\cite{supplement}. Hexagonal tiles, although rare, are also observed with length scale similar to the $P$-tiles [Fig.~\ref{stm}(d)]. The different motifs formed by the congregation of such tiles are highlighted by  circles  in Fig.~\ref{stm}(a). A distinctive motif of the Sn layer that is not observed for $i$-Al-Pd-Mn is a circular congregation   having common sides that we refer to as wheel motif (yellow dashed circles), while half-circular or incomplete wheel motif is named as crown motif  (white dashed circle).
  An expanded view of a crown motif, seen in the bottom right corner of Fig.~\ref{stm}(a), is shown in Fig.~\ref{stm}(e). It shows  five $P$ tiles (C1 to C5) clearly not in the same plane forming a half circle. Noteworthy is that the height  profiles (Fig.~S1(b) of SM\cite{supplement}) along the sides of these tiles show the maximum puckering to be about 0.06~nm, which is sizeable compared to the  thickness of the 1st layer (0.26 nm).  The  puckered nature of the Sn layer can also be quantified by  the root mean square roughness ($S_q$), which turns out to be 0.04~nm, which is twice that of $i$-Al-Pd-Mn ($S_q$= 0.02~nm). A hexagonal tile centered motif  in red circle in the top left side of  Fig.~\ref{stm}(a) also  indicates that the tiles are not in the same plane. It is noteworthy that even in the submonolayer regime, for example at 0.6 ML, as the  Sn adatoms  form islands, the crown motif  is observed [white dashed circle in Fig.~\ref{stm}(f)]. 
In contrast to the $P$-tiles that are uniform, pentagons  with bright vertices with  side length of about 1.1~nm   are occasionally observed [white circles in Fig.~\ref{stm}(a)]. These are different from the white flowers of $i$-Al-Pd-Mn because here a central dark region is observed.

The LEED  of the Sn monolayer is distinctly different from the  substrate:  the latter   recorded with  beam energy ($E_p$) of  81 eV exhibits two sets of ten spots, inner and outer [Fig.~\ref{stm}(g)], in agreement with literature\cite{Gierer_prl97,Barbier02}. A blue pentagon  connects the  intense  (10000) spots of the inner set, while  (00$\bar{1}$$\bar{1}$0) and (11000)  spots  on the outer set are also indicated by circles.   As Sn coverage increases,   the intensities of (000$\bar{1}$0) spots increase  (white dashed pentagon) whereas in contrast the (10000)   intensities decrease (blue pentagon)  [Fig.~\ref{stm}(h)]. Furthermore,  a  set of five innermost spots appears forming a  smaller  pentagon (yellow dashed) with an orientation  rotated $36$$^{\circ}$ $w.r.t$ the blue  pentagon.    In the LEED pattern for 1 ML Sn in Fig.~\ref{stm}(i),  the spots joined by white and yellow dashed pentagons remain clearly visible, while the (10000) spots become inconspicuous. At  other $E_p$ too, the LEED pattern  is modified (Fig.~S2 of SM\cite{supplement}). Consequently, the   intensity versus beam energy (IV) curves for   (10000) in Fig.~\ref{stm}(j) (as well as for  (000$\overline{1}$0), (00$\overline{1}$\,$\overline{1}$0) and (11000)  spots in Fig. S3 of SM\cite{supplement}) are different between the Sn monolayer and  $i$-Al-Pd-Mn, demonstrating the structural differences between the two.

\begin{figure*}[tb] 
	\includegraphics[width=200mm,keepaspectratio]{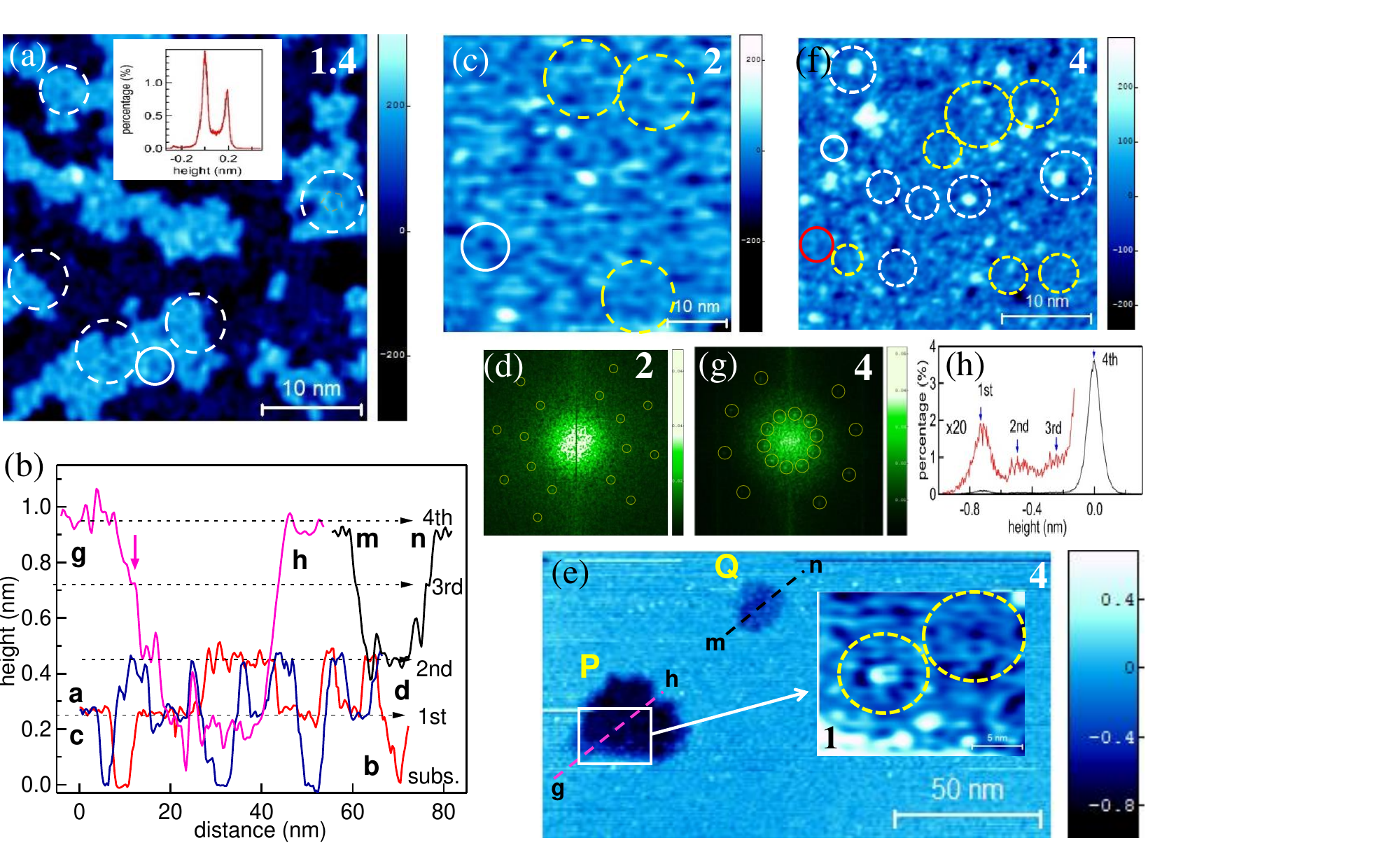} %{Sn_AlPdMn_4ML_fig_vks_ver6_srb_43a}
	% Source: : analysis/STM/Figure/Sn_AlPdMN_4ML_fig_vks_ver6_srb.ppt; Slide 43 saved as pdf using Save --> Pdf --> option --> choose slide number

	\vskip 10mm
	\caption{ %{\bf Scanning tunneling microscopy up to 4th Sn layer}: 
			(a)  High resolution STM topography image  
		~of 1.4 ML Sn/$i$-Al-Pd-Mn with $I_T$= 0.6~nA and $U_T$= -2.6 V  at RT showing motifs that are encircled, the height distribution histogram in the inset shows two peaks corresponding to the 1st and the 2nd Sn layer.   (b)  Height profiles along different directions as indicated in text. 		(c) STM topography image of the 2nd Sn layer at RT, showing different motifs enclosed in dashed (yellow) and solid (white) circles. Coverage in ML is shown in the right hand top corner of all the images.     	(d)  Fourier transform  of   (c) corresponding to the 2nd layer, the spots are highlighted by yellow circles.
			    (e)  STM topography image ($I_T$= 0.3~nA, $U_T$= 2.5~V) of the 4th Sn layer  measured at RT after deposition at  150 K. The  inset (white arrow) is the expanded part of the rectangle  that shows the topography  of the floor of the pit $P$ to be similar to the Sn monolayer in Fig.~\ref{stm}.	 (f) High resolution STM topography image ($I_T$= 0.2~nA, $U_T$= 2.5~V) of the 4th Sn layer showing different motifs enclosed in dashed/solid circles and (g) is the Fourier transform corresponding to the 4th layer.
	 (h) Height histogram of the STM image  in (e). } 
	\label{stmmulti}
\end{figure*}

\subsection{STM and LEED  of Sn multilayers}
The  growth of the 2nd layer is initiated by formation of condensed islands, as shown in the STM image for 1.4 ML Sn/$i$-Al-Pd-Mn [Fig.~\ref{stmmulti}(a)]. The height distribution histogram [inset,  Fig.~\ref{stmmulti}(a)] and the height  profiles given in Fig.~\ref{stmmulti}(b) along $ab$ and $cd$  in an extended area image in  Fig.~S4\cite{supplement}  show that all the islands representing the second layer are $\approx$0.2 nm thick indicating layered growth.  This is expected\cite{Bauer58}   because  the surface energy of Sn (0.71 $J/m^2$)\cite{Tyson77,Vitos98} is less than that of $i$-Al-Pd-Mn (0.82 $J/m^2$)\cite{Dubois06}.    Fig.~\ref{stmmulti}(c)  shows the fully formed  2nd layer, and the motifs are highlighted by circles. The quasiperiodicity  is confirmed by the FT [Fig.~\ref{stmmulti}(d)] that shows two sets of spots (yellow circles) with tenfold symmetry, whose distance from the center is in ratio of $\tau$.  The motifs are  similar to the monolayer of Sn discussed in the previous section. The  wheel  (yellow dashed circle) and pentagon with bright vertices (white circle)  are observed. The second layer is also puckered with $S_q$= 0.05~nm.

In order to enhance the thickness, we have deposited Sn at a lower temperature (150~K) since it is very difficult  to grow thicker layers at RT despite longer time deposition because of  low sticking rate of Sn.  The occurrence of 4th layer is established by analyzing the pit P  in lower left corner of Fig.~\ref{stmmulti}(e). The inset  is the expanded image of the floor of  P that is enclosed by white rectangle. It shows Sn motifs  similar to a  Sn layer and hence  it  is at least 1 ML thick. Thus, the three weak peaks in the   height distribution histogram [Fig.~\ref{stmmulti}(h)] can be assigned to the 1st (floor of the pit P), 2nd (floor of the pit Q), and 3rd layer (at edges of pits P and Q), while the main peak corresponds to the 4th layer. The height profiles along lines $gh$ and $mn$ in Fig.~\ref{stmmulti}(b) also show this, the transition from 3rd to 4th layer is shown by a change in slope (pink arrow). In fact, the pink height profile along $gh$ shows formation of three Sn layers  of nearly equal heights. This is incompatible with the quasiperiodic step heights of the $i$-Al-Pd-Mn substrate, hence confirming the growth of four Sn layers. %Also, if we relate  the pit P to the Sn adlayer, since its depth is $\approx$0.7 nm and the Sn layers  are observed in the wall of the pit in both the height profiles and the histogram  [Fig.~\ref{stmmulti}(b,h)]. In contrast,  the pits in substrate $i$-Al-Pd-Mn  are very rare and even if these are present, should be of depth 0.4~nm, since their terrace step height is 0.4~nm\cite{Sharma06J}.
   
   A high resolution STM image for  the 4th layer in Fig.~\ref{stmmulti}f show the  wheel, crown   (dashed circles),  
    and the triplet motifs (red circle) that are similar to the thinner layers.  FT corresponding to the 4th layer [Fig.~\ref{stmmulti}(f)]  exhibits two set of spots [Fig.~\ref{stmmulti}(g)], whose distance from the center is in the ratio of 1:${\tau}^2$.    The LEED  patterns %,  which represent a macroscopic area of the  surface {\bf in millimeter scale},
    ~ exhibit five fold symmetry within the coherence length scale of the  instrument  (Fig.~S5\cite{supplement}). It shows three sets of sharp spots comprising of an inner set of 9 spots forming a decagon (connected by yellow lines, the 10th spot is hidden by the shadow of the electron gun), 4 intermediate spots forming a pentagon (green dashed lines) and 9 outer  spots (dashed pink lines). Intensity profiles through each of the spots are also shown in Fig.~S5. Quasiperiodicity is reasserted by  the distances of the innermost, inner and outer set of spots from (0, 0) in ratio 1:1.63:2.7 $\approx$ 1:$\tau$:$\tau^2$.  
          Here,  it is worth mentioning that in contrast to the present results, Sn grown on $i$-Al-Cu-Fe showed  island growth after the 1st quasiperiodic layer\cite{Sharma_prb05} possibly because in that work  the deposition was done at much higher temperatures (573-623~K) and large deposition rate of 15~ML/min,  where activated diffusion of Sn across terraces might have resulted in clustering and island growth. In contrast, our deposition temperature for growing thicker layers is 150 K and the  deposition rate is much lower (see section II).      
      
      \begin{figure*}[tb] 
      	\includegraphics[width=165mm,keepaspectratio]{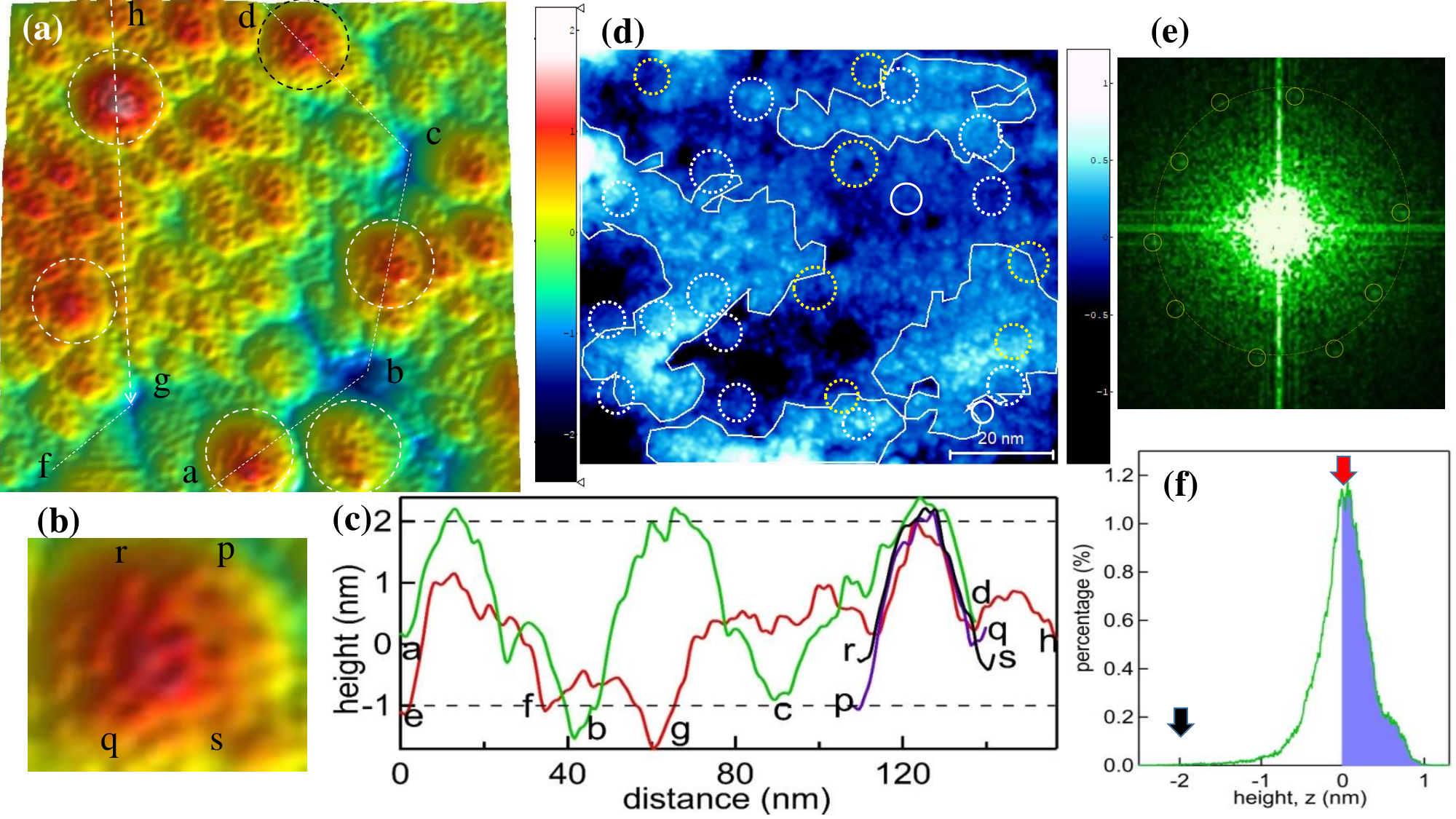} %{3_Sn_AlPdMn_2019_prr_final_figs_ver1_15}%{Sn_AlPdMn_4ML_fig_vks_ver6_srb_44b} 
      	% Source: : analysis/STM/Figure/Sn_AlPdMN_4ML_fig_vks_ver6_srb.ppt; Slide 44 saved as pdf using Save --> Pdf --> option --> choose slide number
      	\vskip 5mm
      	\caption{ %{\bf Scanning tunneling microscopy of Sn thick layer}:  
      		(a) A 104\,nm$\times$106\,nm STM topography image of quasiperiodic Sn showing three dimensional growth   using  $U_T$= -1.5~V, $I_T$= 0.5~nA; measurement at 80~K after deposition at 150~K. (b) An expanded image of a dome enclosed by black dashed circle in  A.  (c) Height profiles along different directions of (a) and  (b), see text.  (d) STM topography image from a different region  showing the motifs that are highlighted. The regions enclosed by white contour lines have thickness $>$3nm.  (e)  The FT of the image in (d) showing ten-fold spots and (f) the height distribution histogram of (d).}
      	\label{dome}
      \end{figure*}

           \begin{figure*}[tb] 
      	\includegraphics[width=165mm,keepaspectratio]{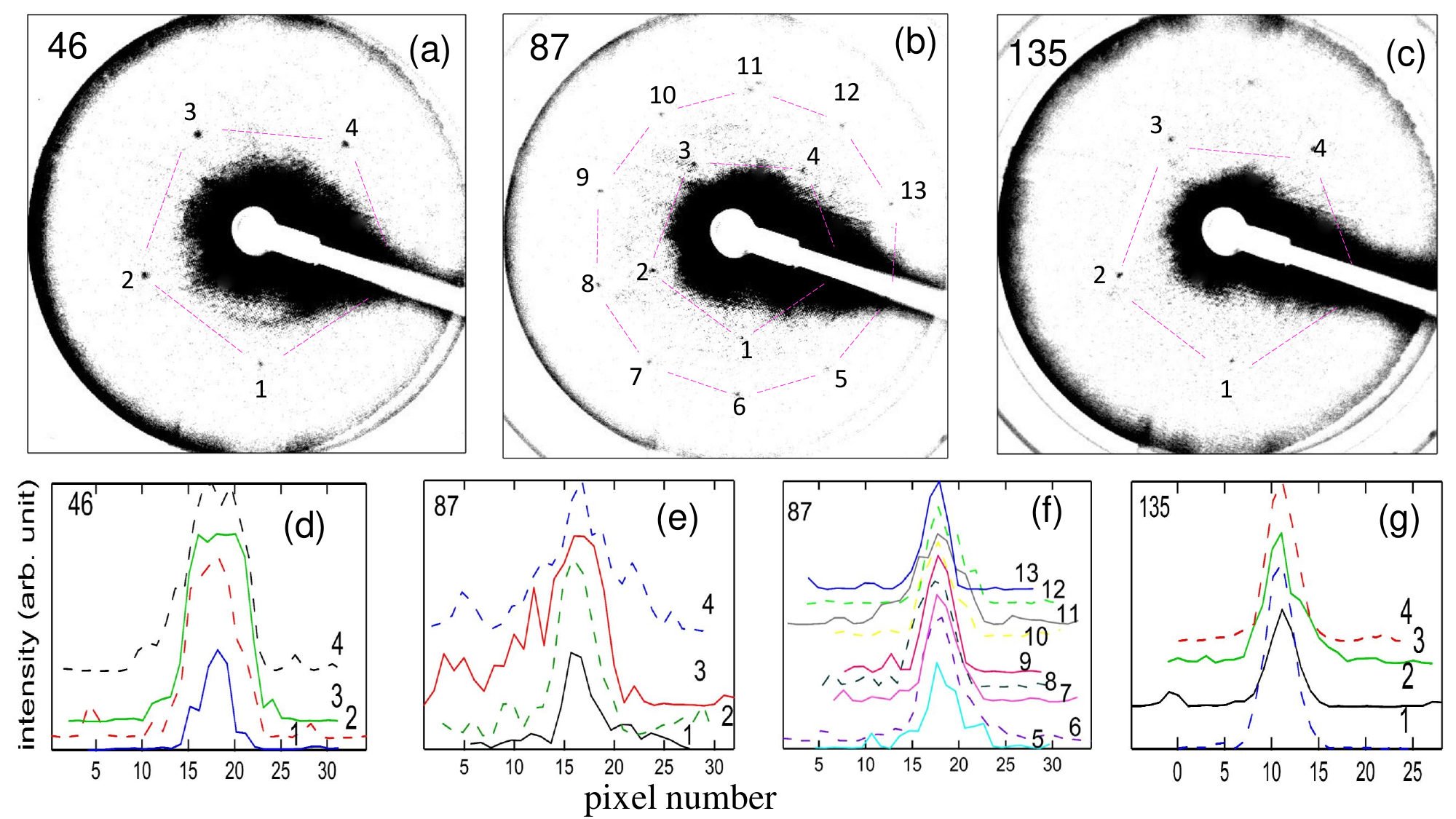}
      	      	\caption{ LEED patterns of the 3-4 nm thick Sn layer with beam energies (a) 46 eV (b) 87 eV and (c) 135 eV. The intensity profiles through the numbered spots (d) 1 through 4 for  beam energy $E_p$= 46~eV, (e, f) 1 through 13 for $E_p$= 87~eV, and (g) 1 through 4 for $E_p$= 135~eV.}
      	      	\label{thickleed}
      \end{figure*}
      
      \subsection{Thick Sn layer studied by STM and LEED}
       In order to obtain quasiperiodic Sn of even larger thickness, the deposition was done at 150 K for twice the time  %where Sn/Pd Auger ratio is found to be about 5 times of the   
   ~compared to the previous (4th layer) deposition and the layer was cooled down to 80~K. In this case, the STM image in Fig.~\ref{dome}(a) shows  dome-like structures  of uniform shape with a  circular base (highlighted by  dashed circles). %The diameter of the circular base is about 20 nm, although smaller domes are also visible. 
~Interestingly,  an expanded view of a dome in Fig.~\ref{dome}(b) shows a central pentagonal tile surrounded  by other pentagonal tiles forming a network. The maximum height of these domes is 3 nm from the base, as shown from the height profiles [Fig.~\ref{dome}(c)] along  $pq$ and $rs$ of Fig.~\ref{dome}(b) and along $abcd$ and $efgh$ of Fig.~\ref{dome}(a).    The diameter of the base of the domes exhibits a Gaussian distribution centered at 17.5~nm with a FWHM of about 10 nm, as shown in Fig.~S6(a)\cite{supplement}.  The puckered nature of the surface is increased by a large amount as portrayed by $S_q$= 0.4 nm. % {\bf  to MM: can we relate them to the clathrate cages? But the dimension is too large, I guess!}  {\bf MMnote: I think such broad Gaussian distribution of the dome bases is not directly related to the characteristic lengths in clathrate structure.}

However, possibly due to a competition between  three dimensional   versus layered growth, the domes are not visible everywhere on the surface. For example,  in Fig.~\ref{dome}(d), a different region shows relatively longer range undulations of $\approx$50 nm (see also Figs.~S6(b) and S6(c)\cite{supplement} for an extended area image and height profile). It is relatively flat compared to Fig.~\ref{dome}(a)  with smaller $S_q$= 0.3 nm. Here, motifs such as wheel (yellow dashed circle), crown (white dashed circle), pentagon with bright vertices (white circle)  are  observed. Moreover, in Fig.~\ref{dome}(e), the ten fold FT confirms the quasiperiodic nature of the layer; a marginal deviation in the arrangement of these spots from circle could be because of finite amount of thermal drift during scanning.  
  
  In order to find the thickness of this layer, we note that the peak of the height histogram at $z$=\,0 nm (red arrow in Fig.~\ref{dome}f) shows that about 50\% of the area (shaded with  blue color) has a thickness (or height) of $\geq$2 nm $w.r.t.$ the minimum at $z$$\approx$\,-2 nm (black arrow). The minimum region is identified to be around $a$ in Fig.~S6(b) from the height profile $abc$ shown in Fig.~S6(c). Since we obtain a thickness of 1~nm for the wetting layer $i.e.$ up to the 4th layer, an estimate of the thickness of the minimum region around $a$ in Fig.~S6(b) is $\approx$1 nm.  Thus, the  thickness of the regions enclosed by white contour lines in Fig.~\ref{dome}(d) is  $\geq$3~nm. Presence of many quasiperiodic motifs (encircled) in this region that  are similar to the motifs observed in the 2-3 nm thick region outside the contour lines  or those in Figs.~\ref{stm},~\ref{stmmulti}  indicates similar quasiperiodic structure in the $\geq$3~nm region.  An estimate of the height of the Sn  domes   in Fig.~\ref{dome}(a), by the same argument discussed above, is $\geq$4~nm, since these are at a height of 3~nm from the base (Fig.~\ref{dome}(c)), and the base can be taken to be $\geq$1~nm in thickness.
         The appearance of domes and undulations [Fig.~\ref{dome}(a,d)] is indicative of transformation towards quasiperiodic ordering in the vertical direction, which could be expected as the Sn-adlayer becomes thicker. The growth  transforms from puckered but primarily two dimensional to  quasiperiodic growth in all three directions similar to a bulk quasicrystal.  We  find that the thick film is not stable when warmed up to RT, as is evident  from  40-50\%  decrease of the Sn Auger signal, which is possibly caused by the Sn atoms diffusing out to the sides of the substrate. Diffusion of Sn from the back of Al-Ni-Co substrate to the front with increasing temperature has been reported in literature\cite{Shimoda_jns04}.

  Further evidence of  quasiperiodicity in this thick layer in the macroscopic length scale is provided by the LEED patterns %with energies of 46, 87 and 135 eV  
  ~that show sharp five fold spots [Fig.~\ref{thickleed}(a-c)]. The  intensity profiles of each of the spots are shown in [Fig.~\ref{thickleed}(d-g)].   At 87 eV [Fig.~\ref{thickleed}(b)], the outer 10 spots and the inner pentagon are observed and the ratio of their distance from the (0,0) specular spot is 1:1.63 $\approx$ 1:$\tau$.  At beam energies of 46 eV and 135 eV [Fig.~\ref{thickleed}(a,c)],  four spots representing a pentagon are observed,  the fifth spot is hidden by the shadow of the electron gun. In Fig.~S7\cite{supplement},  the IV curves for the thicker Sn layers are shown, their peak positions are similar to that of 1 ML Sn, indicating their structural similarity. LEED patterns as a function of beam energy are shown by  the video files in SM, such as,  {\it 3nmleed} for the 3 nm Sn layer,    {\it 1.8nmleed} for 1.8~nm, {\it 1nmleed} for  1~nm  and  {\it 0.25nmleed} for the 0.25~nm (or 1 ML) Sn layer thickness\cite{supplement}.
  If an adlayer has different domains,  it results in the appearance of multiple spots in the LEED pattern. There are reports of such rotational epitaxy for different metals  grown on $i$-Al-Pd-Mn\cite{Sharma,Bolliger}. On the contrary, our LEED patterns (Figs.~\ref{stm}, \ref{thickleed}, S2, S5 and the video files) show that all the  spots move towards the (0,0) spot with increasing beam energy. Importantly,  no extra spots appear or splitting of the spots occur over the whole beam energy range.  We also do not observe any signature of spots arranged in a periodic fashion that is characteristic of crystalline or approximant phases. 
  	%  	 These videos show that all the LEED spots move towards the (0,0) spot with increasing beam energy and also  no extra spots appear or splitting of the spots occurs over the whole beam energy range. This rules out existence of crystalline domains of the adlayer with a specific orientational relationship with the substrate, since in that case multiple spots %and splitting of spots with beam energy ~ are observed.\cite{Sharma,Bolliger} % There are reports of such rotational epitaxy for different metals  grown on $i$-Al-Pd-Mn . %We do not find any such evidence   from our LEED patterns (Figs. 1,  4, S2, and S5 and the video files). 
    ~We however note that the thicker layers have disorder, as also corroborated by the theoretical clathrate model that is discussed later, which % The roughness increases with thickness, as shown by the calculated root mean square roughness (Sq) from the STM images. 
  ~results in relative weakening of the LEED spots at larger thickness. But, what is very important is that all the LEED patterns show only quasiperiodic spots with no hint of any alternative ordering.
  
%   demonstrate movement of the spots towards the (0,0) spot;  five fold quasiperiodicity is reconfirmed by the absence of any splitting of the spots and appearance of extra spots over the whole beam energy range.  %{\bf ~Another confirmation of  the structural similarity  is that the positions of the LEED spots relative to the (0,0) spot remain unchanged over the whole coverage range, as shown in Fig.~S8. % by the intensity profiles through the spots numbered 2, 8 and 5 for different coverages up to 5 ML. 
 % ~{\bf However, the  quality of the LEED pattern worsens and the intensities of the IV curves  are reduced as thickness increases due to increase of disorder and defects (Fig. S7).} 

\subsection{Quasicrystal clathrate model of Sn}

%%%%%%%%%  bulk clathrate energies relative to Diamond
% Si clath +0.053 (cF136) [eV/atom]
% Ge clath +0.026 (cF136) [eV/atom]
% Sn clath +0.023 (cF136) +0.028 (cP46) [eV/atom]

%We note that direct modeling of the Sn adlayer on $i$-Al-Pd-Mn quasicrystal is a challenge due to the chemical and structural complexity of the combined system, implying a large system size and the necessity of \textit{ab-initio} approach. A more  pragmatic strategy would be to conceptualize an alternative monoatomic Sn structure that would be compatible with the $i$-Al-Pd-Mn surface. 

The motifs observed from scanning tunneling microscopy as well as the low energy electron diffraction patterns show structural signatures for the Sn film that are distinct from the substrate. Sn thus exhibits intrinsic quasiperiodic growth to a thickness of 3-4 nm, which almost in the realm of bulk-like quasiperiodic growth where impact of the substrate becomes negligible. This is because the effective potential of  a solid including the exchange interaction term perpendicular to the surface is almost zero beyond unity Fermi wavelength\cite{Lang70}. Al and Mn that form the top layer of $i$-Al-Pd-Mn both have Fermi wavelengths of about 0.36~nm,% and 0.37~nm, respectively 
~which is almost an order of magnitude less than the 4 nm thickness of the quasiperiodic  Sn layer.  The function of the   substrate is  to suppress initial formation of stable crystalline forms of Sn $i.e.$ the $\alpha$  and $\beta$-Sn.

\begin{figure*}[tb] 
	\includegraphics[width=130mm,keepaspectratio]{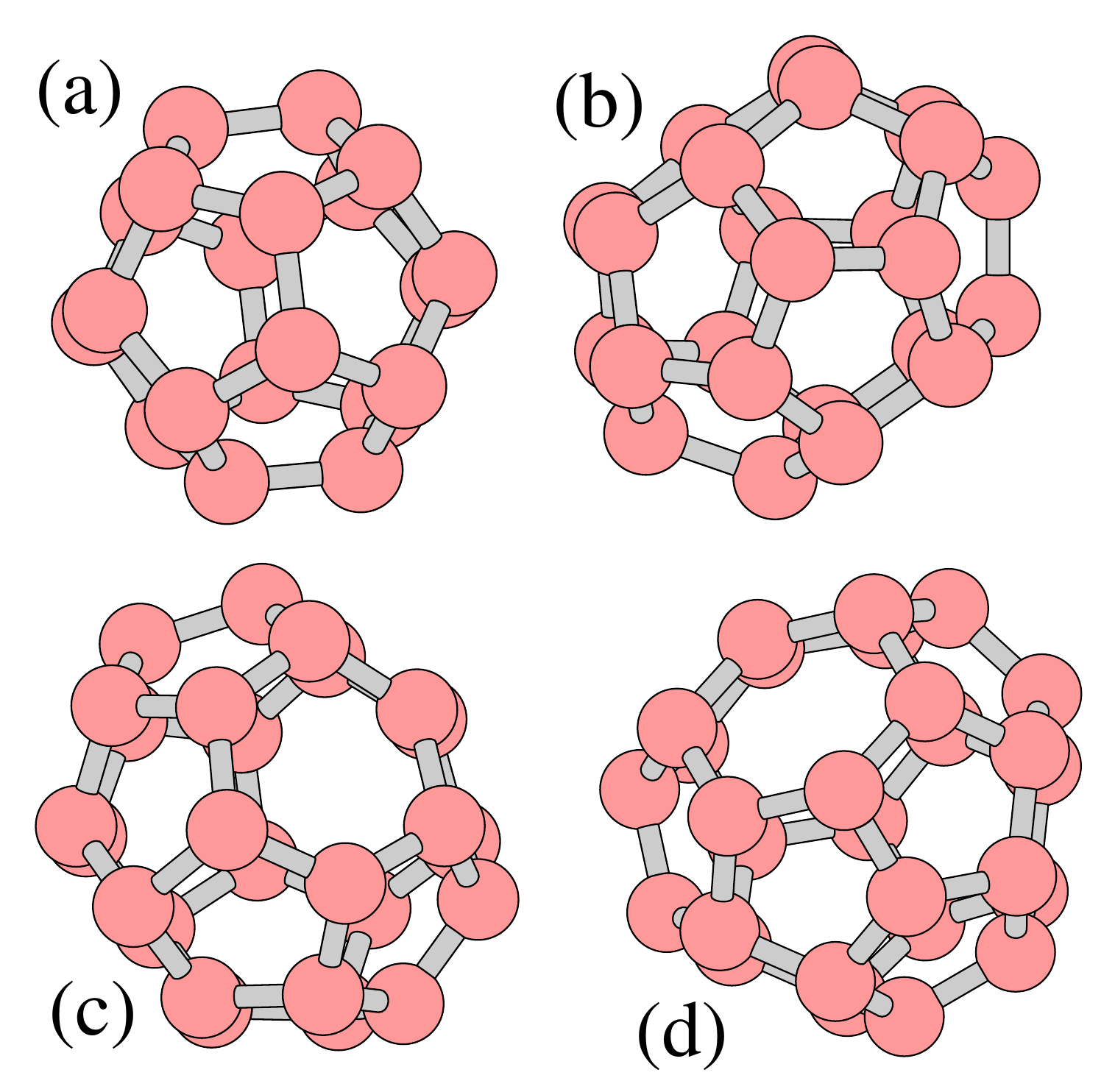} %{Final_figures_Sn_AlPdMn_2019_ver8_fig2}
	%	\vskip -10mm
	\caption{ The four different types of clathrate cages: (a)		
		The smallest cage is a dodecahedron with coordination (CN) of 20. It is bounded
		exclusively by pentagonal faces and has full icosahedral symmetry. The
		bigger cages are bounded by either pentagonal or hexagonal facets with
		(b) CN= 24, (c) CN= 26  and (d) the largest with CN= 28.}% The diamond structure of the $\alpha$ phase contains exclusively hexagonal planes. 
			
	\label{cages}
\end{figure*}

 In the Sections~III.D through III.I, we combine the geometrical concepts and DFT calculations, while in Section~III.J,  we discuss how  Sn white flower (SnWF) cluster facilitates nucleation of the intrinsic Sn structure.  We show that  the intrinsic Sn structure is a metastable clathrate quasicrystal, and that in a narrow window of covering widths of 3-4 nm, an unsupported pentagonal clathrate structure is more stable compared to  $\alpha$-Sn. In the $\alpha$-Sn structure, atoms are tetrahedrally coordinated and satisfy the $sp^3$ type bonding scheme.   But there exists a whole family of alternative structures -- {\it clathrates} -- satisfying the same tetrahedral bonding scheme  (Fig.\ref{cages}), achieving that goal by grouping atoms around point centers into empty ``cages''. %{\bf The four different clathrate structures considered by us are shown in Fig.\ref{cages}.} 
 ~Indeed ``empty'' clathrates are experimentally achievable metastable states of Ge\cite{Guloy} or Si\cite{Gryko}, and there is a broad family of clathrate structures that are stabilized by addition of large ``host'' atoms captured inside the cages. The clathrate structures are modular in the sense that the constituting polyhedral cages can be nontrivially recomposed into variety of periodic arrangements.

\begin{figure*}[tb] 
	\includegraphics[width=170mm,keepaspectratio]{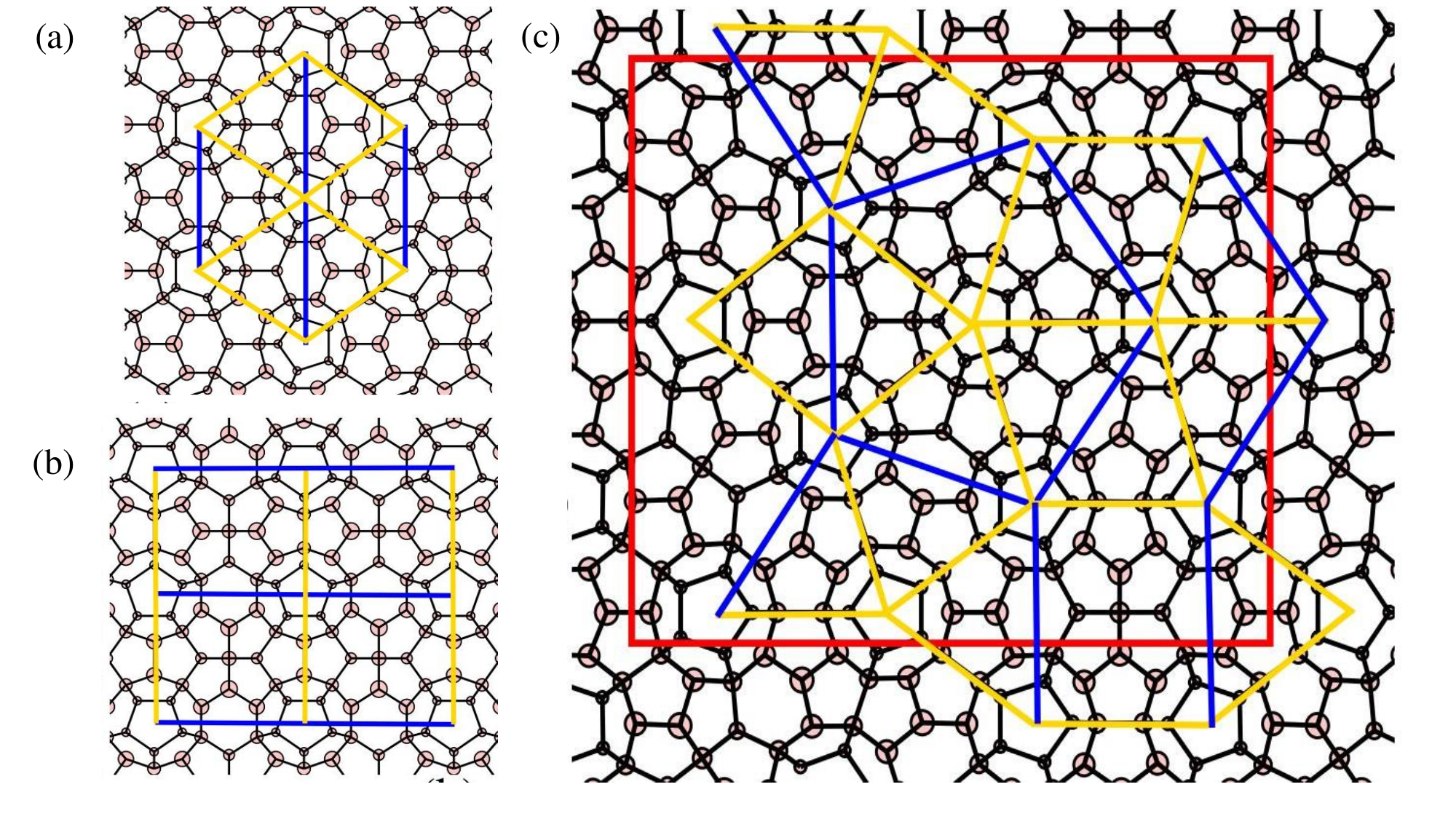} %{Final_figures_Sn_AlPdMn_2019_ver8_fig2}
	%	\vskip -10mm
	\caption{ %{\bf  clathrate model}: 
		6\AA~ thick sections through atomic structures of three clathrate approximants: (a) 136-atom cubic structure type II viewed along (110) direction; (b) 40-atom type III hexagonal structure viewed along (100) direction, and (c) 392-atom orthorhombic clathrate serving as ansatz for the decagonal clathrate. All the structures are shown along their common pseudo-decagonal axis. The decagonal (in projection) rings can be associated with the vertices of tiling of rectangle- triangle (R-T) tiling. Case (a) is pure triangle tiling, case (b) is pure rectangles. Models are constructed by dual transformation of Frank Kasper structures: Cu$_2$Mg (a) Al$_3$Zr$_4$ (b) and hypothetical 22-atom orthorhombic decagonal approximant (c). The blue linkages are 12.6\AA~ long.} 
	
	\label{clathrate}
\end{figure*}

Two of the known clathrate structures commonly designated as type II and III exhibit special relationship, illustrated in  Fig.~\ref{clathrate}(a,b) -- they share a common pseudo 5-fold axis, and equal vertical periodic repeat that is  about 12.7\AA~ in case of Sn. Size of the circles representing the atoms scales with their vertical height, and the neighboring atoms are connected by black lines along  2.9\AA~ long interatomic bonds. Associating the centers of the decagonal rings	(decorated inside by pentagons) with the tiling vertices reveals the geometric backbone of the structures:  tiling of isosceles triangles T with wider angle 72$^{\circ}$ (two yellow and the longer blue sides), and rectangles R with aspect ratio	$\sqrt{\tau+2}$/$\tau$; blue linkages denoted $b$  are about 12.6\AA~ long, and the shorter yellow linkage $y$ is given by  $b\tau/\sqrt{\tau+2}\sim$10.82\AA. Since the tile decoration is symmetric and identical for all the yellow edges in both structures, and the same holds for the blue linkages. The cage motifs from two prototype structures can be combined to form a family of variant structures that are in one-to-one correspondence with planar rectangle--triangle (R-T) tilings. The R-T tilings may be arranged periodically as in Fig.~\ref{clathrate}(c), but the local 5-fold axial symmetry of the cages eventually leads to global pentagonal order of the quasiperiodic structures\cite{COC}. %or “random” tilings.\cite{OZ} 
~Thus the clathrate recipe for R-T tiling decoration gives rise to whole family of pentagonal structures, periodic, quasiperiodic or disordered.
 
 We note that the  cage-cage linkage length in Sn clathrates of $\approx$12.6\AA~ almost exactly coincides (to within 1\%) with the pseudo-Mackay cluster-cluster separations in $i$-Al-Pd-Mn of 12.55\AA~ ($\tau$-times the intercluster linkages of 7.7\AA~ between ``mini-Bergman'' or between ``pseudo-Mackay icosahedra'' clusters). In the clathrates, the prevalent dodecahedral cages have icosahedral symmetry, and sections through the cage normal to a 5--fold axis will have pentagonal symmetry. Since the symmetry of the LEED patterns [Fig.~\ref{thickleed}, Fig.~S5] is determined by the  surface layer, we predict that in general their symmetry will be 5-fold. 

\subsection{Quasicrystal clathrates as duals of the Frank-Kasper structures}

In Frank--Kasper (FK) structures, the space is divided into face-sharing tetrahedra. Decorating each tetrahedron center by Sn atom, every tetrahedron's face is intersected by a bond between pair of neighboring Sn atoms. Since there are four faces for every tetrahedron, every Sn  atom is guaranteed to have four neighbors in the directions of the surrounding face centers. Thus, tetrahedrally close packed structures are in one to one correspondence with their dual, clathrate relatives\cite{FK, FK59, OAS}. Four canonical FK polyhedra with triangulated coordination shells -- Z12, Z14, Z15 and Z16 -- become respectively 20, 24, 26 and 28-vertex cages, whose faces are pentagons and hexagons. This dual relationship is an alternative path to the decagonal clathrates defined above, by dual-transforming decagonal FK structures. The two clathrates from Fig.~\ref{clathrate}(a, b) are in fact dual relatives of the FK prototypes  Cu$_2$Mg and Al$_3$Zr$_4$ FK structures. Moreover, we could generalize clathrates to include 12-fold dodecagonal quasicrystals based on FK decoration of square-triangle tilings\cite{II}, whose strictly quasiperiodic version would be based on Stampfli’s square-triangle inflation rule\cite{ZU}.

The possibility of icosahedral FK structure whose dual would be perfect icosahedral clathrate is not clear. The only locally icosahedral and strictly tetrahedrally close-packed structure known is, to our knowledge, the ``1/1'' icosahedral approximant with prototypical natural realization in Al-Mg-Zn system, discovered by Bergman $et~al.$\cite{Bergman52}  In their milestone paper, Henley and Elser\cite{HE} described this structure as a simple decoration of Penrose rhombohedra, and proposed that generalization of that decoration scheme eventually leads to FK quasicrystal enjoying global icosahedral symmetry. But there is a bottleneck: the decoration rule requires that all “oblate” rhombuses in the 3D Penrose tiling are paired along their short body diagonals, but such tiling of 3D golden rhombuses is not known at present. An interesting aspect of the bulk icosahedral clathrate is its possible genuine entropic advantage over layered quasicrystals.

Relaxing the strict $sp^3$-bonding requirement (that translates into tetrahedral bonding rule) brings interesting clue between spontaneously formed icosahedral quasicrystals in toy mono-particle system interacting via oscillating pair potentials\cite{qcla} and real tin-based systems. As is clear from the phase diagram outlined in Fig.~3 of Ref.~\onlinecite{qcla}, the ``low-density'' icosahedral phase neighbors with clathrates, and major coordination number of that structure is four, although coordinations up to seven may occur. Given that Sn character is intermediate between $sp^3$-bonding type and metallic, which is reflected by the close competition of $\alpha$ and $\beta$ tin variants, such ``imperfect'' quasicrystal ordering seems entirely plausible.

\subsection{Cohesive energies and diffraction patterns of bulk Sn-clathrate models }

\begin{table}
	\begin{tabular}{|c|c|c|c|c|c|c|c|c|} \hline
		TYPE   &     FK prototype  & name & C20 & C24 & C26 & C28  & N$_{at}$ &   $dE$  \\ %   # Z12 Z14 Z15 Z16 #  dE(Ge)
		\hline
		II          & Cu$_2$Mg.cF24                & Sn.cF136    &0.666 & -- & -- & 0.333 &  34$^a$   &  28  \\  %  
		III         & Al$_3$Zr$_4$.hP7               &  Sn.hP40 & 0.429 & 0.286 & 0.286 & --  &    40        &  40   \\
		decagonal   & Mg28Zn41.oP69$^{b}$ & Sn.oP392 &0.594 &0.087  & 0.087 & 0.232   &   392$^a$  &  34   \\
		icosahedral & AlMgZn.cI162$^{c}$ & Sn.cI1840   &   0.593 & 0.259 & 0.074 & 0.074 &920$^a$  &  48   \\  % 36+32 8 8 28 ; fractions 0.61 0.07 0.07 0.25
		\hline
	\end{tabular}
	~~\\
	~~\\
	\noindent $^a$ calculation in primitive cell; 
	$^{b}$ hypothetical structure;
	 $^{c}$ in the actual intermetallic structure of AlMgZn, two atoms -- center of Bergman clusters -- are vacant\\
	% REFERENCE VOLUME OF FCC DIAMOND IS 36.57A^3/atom. Clathrates are about 41.4, by 13.3% larger.
	\caption{ \label{tab:struc}
		Bulk DFT energies ($dE$, in meV/atom) of the Sn clathrates,
		relative to $\alpha$-Sn.cF8. First two rows are the basic clathrate
		types corresponding to decorated isosceles triangle tiles [type II, shown in Fig.~\ref{clathrate}(a)] and rectangles [type III, Fig.~\ref{clathrate}(b)] . Decagonal approximant (third row) combines cage motifs from types II and III structures following 
		rectangle-triangle tiling geometry [Fig.~\ref{clathrate}(c)]. Last row reports 1/1 icosahedral approximant. Second  and third columns show prototype and Pearson symbol of a dual FK structure, respectively, while 4th to 7th columns  show the fraction of each
		kind of cage in the respective structure (Fig.~\ref{cages}), which are duals to the FK coordination polyhedra Z12- Z16. N$_{at}$ reports the number of atoms per periodic cell.}
%	\lable{table}
\end{table}

While intermetallic clathrates are usually stabilized by 15\% content of large guest atom that fits into large clathrate cages (and do not fit inside $\alpha$-Sn structure), even empty clathrate structures are energetically competitive. Fully relaxed cohesive  DFT energies of these structures relative to the $\alpha$-Sn  structure ground state are shown in Table I.  The smallness of these energy differences (28-48 meV/atom range) as well as the fact that the decagonal approximant energy is intermediate between energies of type II and III prototype clathrates, are the first argument supporting the plausibility of the clathrate structure for Sn.
%{\bf ** SRB: MM please modify/integrate this error related discussion (it is include in the Response) with previous paragraph**:
~We expect that the leading term in energetics of any clathrate structure will depend on the fraction of each type of the cage in the structure. Since the decagonal clathrate approximant (row 3 in Table ~\ref{tab:struc}) cage fractions are intermediate between the two basic approximants (rows 1 and 2), its intermediate energy supports our expectation. On the other hand, the fact that the icosahedral approximant (row 4 in Table ~\ref{tab:struc}) has highest energy, despite the fact that it is also structurally intermediate, suggests that the effective Hamiltonian should also tackle some pairwise cage-cage terms that did not occur in the 392-atom 2D approximant. Nevertheless, we would expect that the optimal bulk quasicrystal energy should lie in between 28 and 40 meV bounds provided by the extremes of fractional cage contents in the two basic type II and III clathrate approximants, but most likely very close to the energy of the decagonal approximant. % Further insights into quasicrystal clathrate energetics should definitely include entropy of a thin slab, and also role of entropy for adding extra layer on the previously grown slab.}

Fig.~\ref{diffraction}(a,b) shows that diffraction patterns of  decagonal Sn-clathrate and the bulk $i$-Al-Pd-Mn phases parallel to their common 5-fold axis exhibit striking similarity in both positions and intensity of the strong diffraction peaks. The apparent symmetry of the diffraction pattern is 10-fold in both cases, due to the presence of inversion symmetry in $i$-Al-Pd-Mn, and 10-fold screw axis
in $d$-clathrate. The similarity of the patterns supports the Sn clathrate structural hypothesis.

\subsection{Energetics of unsupported Sn slabs}

We further test the clathrate structural hypothesis by comparing the cohesive energies of the unsupported Sn slabs. We constructed reference series of the slabs of $\alpha$-Sn %takes its diamond structure 
~with the surface normal to (111) direction, and of $\beta$-Sn with the surface normal to (100). These surfaces are flat planes in both cases, in $\alpha$-Sn atoms on the surface are 3-fold coordinated, in $\beta$-Sn it is 5-fold coordinated (when we cut off the first coordination shell at about 3.3\AA).

For clathrates, we require that all the surface atoms have coordination equal to 3, like $\alpha$-Sn. This can be realized by growing the clathrate layer in discrete steps, by completing all (open) cages whose centers are nearest to the current surface. This strategy allows us to circumvent the difficulty of growing  the ``sparse'' clathrate structure that does not contain well defined flat layers, but rather coalesces sharing pentagonal/hexagonal faces between the neighboring cages.
% This follows from the fact that the dual polyhedra to the cages– FK polyhedra– have exclusively triangulated shells. If the 2-fold and 5-fold coordinations cost prohibitive energy, the cages may be viewed as molecules, that coalesce together by saturating the missing fourth bonds. The clathrate cages then close-pack by sharing the pentagonal and hexagonal faces of the cages- ”molecules”.  

\begin{figure*}[tb] 
	\includegraphics[width=185mm,keepaspectratio]{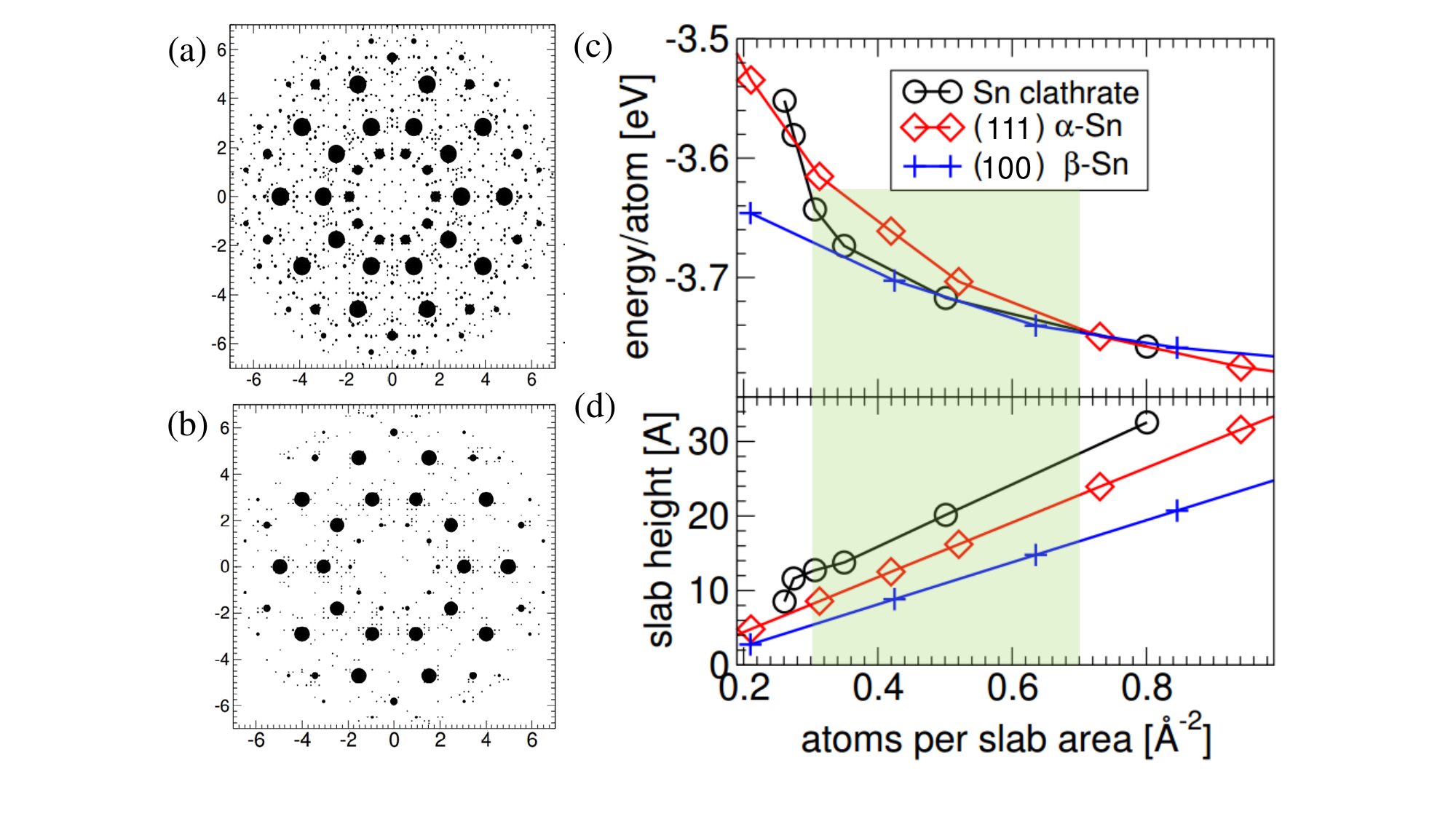} %{Final_figures_Sn_AlPdMn_2019_ver8_fig2}
	%	\vskip -20mm
	\caption{ %{\bf  clathrate model}: 
		The calculated diffraction patterns of the (a) large 2692-atom decagonal clathrate approximant  and (b) $i$-Al-Pd-Mn represented by its “5/3” approximant. (c) Energy  and (d) slab height  as a function of the number of Sn atoms per slab area for  $\alpha$-Sn, $\beta$-Sn and the decagonal  clathrate showing that in the 0.3-0.7 atom/\AA$^2$ range (shaded light green), the clathrate is  stable $wrt$ $\alpha$-Sn. %, whereas at 0.5 atom/\AA$^2$ or 2 nm slab thickness (green arrow) it is comparable to $\beta$-Sn. 
		~The increased height of the clathrate versus $\alpha$-Sn is due to higher density of the latter, and large fluctuation of height on completed surface of the clathrate.} 
	
	\label{diffraction}
\end{figure*}

In decagonal clathrates, the cage centers are located on four heights per 12.7\AA~ vertical repeat: dodecahedra decorate the R-T tiling vertices at fractional height coordinates $z$= 0 and 0.5. The mid-edge and tile-interiors cage centers are all located at $z$= 0.25 and $z$= 0.75. From this we derived the aforementioned strategy for the slab construction: the surfaces are formed by completing all cages with next higher fractional coordinate $z$.

To represent the slab of the decagonal clathrate, we chose periodic approximant from Fig.~\ref{clathrate}(c).
Slabs for all three structures, $\alpha$, $\beta$ and decagonal clathrate approximant were fully optimized by relaxing ionic positions and lattice parameters until the average DFT force was less than 0.05 eV/\AA, followed by single final self-consistent iteration with fixed geometry.

%{\bf MMedit:}
Energetics of sufficiently thick layers are characterized by bulk and surface energies, $E_{surf}= (N \times E_{bulk}-E_{lay})/A/2$, where $E_{bulk}$ is bulk energy per Sn atom, $E_{lay}$ is energy of the layer consisting of N atoms per periodic box, and $A$ is the surface area. The bulk energy of the clathrate Sn approximant is stable compared to that of $\beta$-Sn  by  11 meV/atom. In contrast, the former is unstable against $\alpha$-Sn by 35 meV/atom. %SRB: changed into two sentences since your sentence was complicated. Pls see if the meaning is retained. I also use the stability of clathrate wrt beta Sn in the conlcusion. 
  %, and $\beta$-Sn bulk energy was by another 11 meV/atom less stable than the selected clathrate approximant. 
~The surface energies, on the other hand show opposite trend, reaching minimum for $\beta$-Sn slabs, with $E_{surf}\sim$0.025 meV/$\AA^2$, followed closely by clathrate (0.27 meV/\AA$^2$), and $\alpha$-Sn (0.41 meV/\AA$^2$). This result is consistent with $\alpha$ and $\beta$ surface energies calculated in Ref.~\onlinecite{Hormann15}, leading to prediction that sufficiently small Sn nano-particles will consist of $\beta$-Sn even in the low-temperature limit (although that work does not include the possibility of nanoparticles based on clathrate cages).

In order to capture energetic competition between the three families of slabs, we evaluate and plot DFT energies of the slabs as a function of the number of atoms per unit area normal to the slab surfaces. The results are shown in Fig.~\ref{diffraction}(c), which plots cohesive energy/atom as a function of the area coverage. Fig.~\ref{diffraction}(d)  shows the height defined as a difference of the topmost and bottom atom coordinates ($z_{max}$- $z_{min}$). 

For thinnest slabs with coverages up to about 0.4 atom/\AA$^2$, the most stable structure is $\beta$-Sn. The clathrate approximant performs excellently for layer thickness exceeding approximately its one periodic repeat (about 12.7\AA), or twice the 5-fold diameter of the fundamental dodecahedral cage: in the 0.3-0.7 atom/\AA$^2$ covering range, $i.e.$ for slab heights between 10-35\AA, it is more stable than $\alpha$-Sn [the range is shaded in light green in Fig.~\ref{diffraction}(c,d)]. It approximately equals energy of $\beta$-Sn in the range  0.5-0.7 atom/\AA$^2$ or 2-3 nm slab thickness. %SRB: this is included to make consistent to what you mention in Abstract.** 
~Finally, when the coverage exceeds 0.7 atoms/\AA$^2$ or slab thickness $\geq$ 3~nm $i.e$ approaching the bulk limit, %SRB: include the phrase "bulk limit" to connect to earlier result that in bulk alpha-Sn is most stable
~ the most stable structure is $\alpha$-Sn.

\subsection{Photoemission spectroscopy and  density of states from the clathrate model}

\begin{figure*}[t]
	\includegraphics[width=175mm,keepaspectratio]{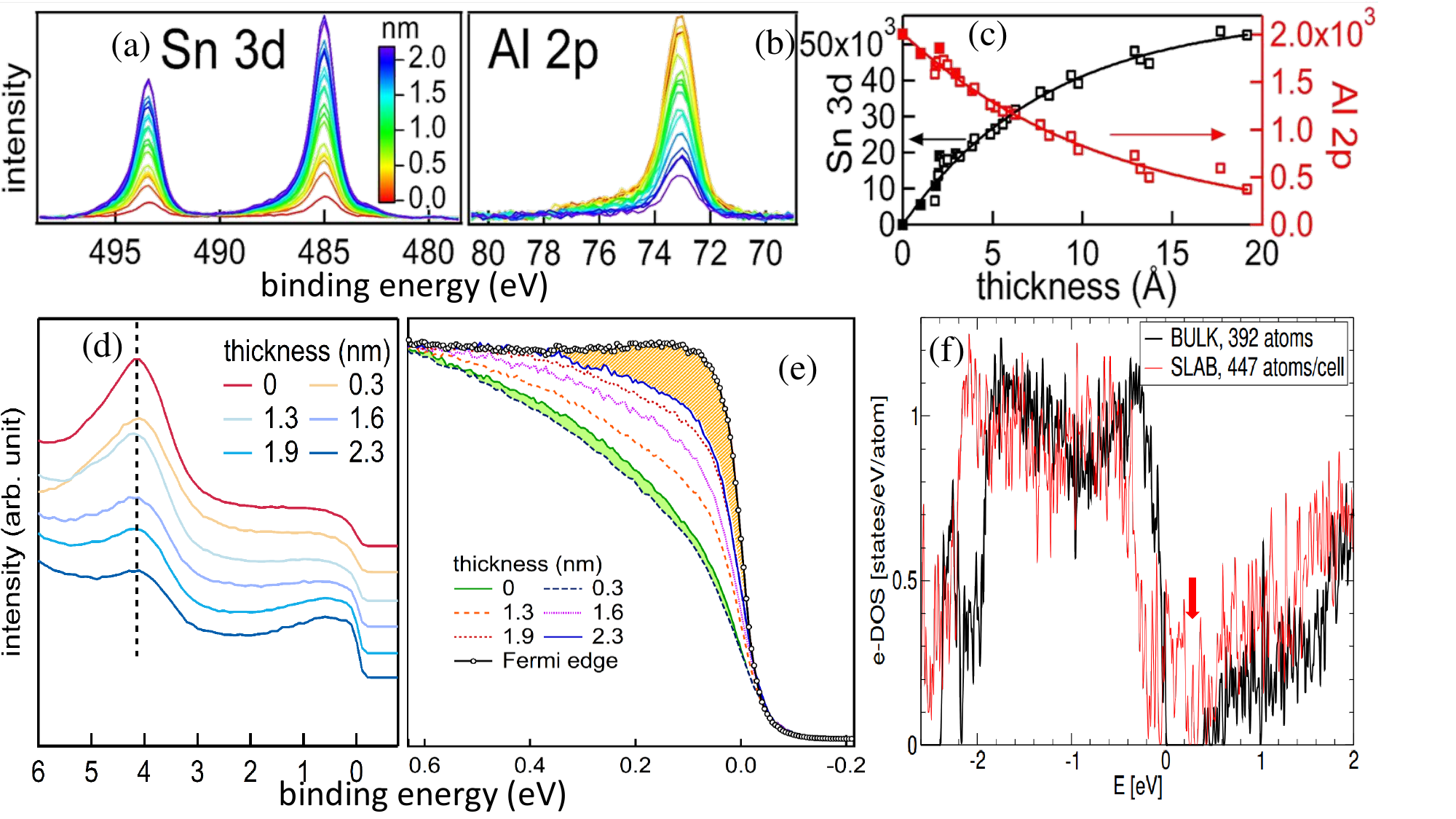} %{7_Sn_AlPdMn_2019_prr_final_figs_ver1_06} %{Layout1_6_Figure5_ver9} 
	%Source: layout1_6 of Figure3_combination_srb_ps_ss.pxp
	%\vskip -10mm
	%{Layout1_4_fig5}%{Layout1_3.pdf}
	% Source: Layout1_3 of analysis/XPS/Figure3_combination.pxp
	
	\caption{ %{\bf Photoemission spectroscopy and  density functional theory:} 
		(a) Sn 3$d$, (b) Al 2$p$ XPS core-level spectra as a function of coverage and   (c)  the  areas of the  core-level peaks  at 100~K 
		~(open squares) and RT (filled squares). The continuous lines are the least square  fitted curves. % showing an exponential variation.
		~ (d) The valence band photoemission  spectra of  Sn/$i$-Al-Pd-Mn using UPS for different Sn coverages, the spectra are staggered along the vertical axis. 
		~(e) The UPS spectral shape close to the Fermi level ($E_F$) for  Sn layers up to thickness of 2.3 nm deposited at 150~K along with the Fermi edge of Mo foil in electrical contact with the specimen. The shaded orange region shows the suppression of intensity of the  thick layer $w.r.t$ the Fermi edge, confirming the existence of pseudogap. The shaded green region shows the deepening of the pseudogap compared to $i$-Al-Pd-Mn. (f) The calculated electronic density of states (DOS) of the 14\AA~ thick relaxed Sn clathrate  slab and its bulk parent structure. The bulk structure is repeated with 12.6\AA~ period, while the slab implementation contains about 14\AA~ thick layer  separated from its periodic image along the surface normal by about 33\AA~ thick vacuum gap.} %Thus,  the slab thickness of 16A exceeds by about 30\% the vertical period of the bulk clathrate and so the slab structure contains correspondingly more independent atoms, than the bulk. }
	\label{ups}
\end{figure*}

 In Fig.~\ref{ups}(a,b),  Sn 3$d_{5/2}$ and  Al 2$p$ XPS core level spectra that  appear  at 485 eV and 73.1 eV,  respectively, show no change in their binding energies with Sn coverage. This is a significant observation because core-level binding energies are sensitive to the transfer of  electronic charge from the adsorbate  to the substrate, which obviously can be ruled out in this case. Absence of core level shift or change in the shape also excludes the possibility of any alloying or intermixing of Sn with the substrate.   This observation  also implies a propensity for condensed island growth\cite{Shukla_prb06} as also shown in Fig.~\ref{stmmulti}(a),  indicating the dominance of   adsorbate-adsorbate interaction.  The Sn 3$d_{5/2}$ core level binding energy (485 eV) is similar to that of $\alpha$-Sn\cite{Hegde},  indicating the similarity of the quasiperiodic Sn layer with that of $\alpha$-Sn, thus reaffirming  the  clathrate model where $sp^3$ hybridization dominates. In Fig.~\ref{ups}(c), the  intensity variation of the Sn 3$d$ and Al 2$p$ XPS core-level spectra  exhibits exponential  trend as a function of  thickness  indicating a layered growth, as also shown by STM. Fig.~\ref{ups}(d) depicts the  valence band of Sn/$i$-Al-Pd-Mn, where the prominent peak at 4~eV binding energy arises from  Pd 4$d$-like states. It decreases in intensity with Sn deposition, but its binding energy  remains unchanged  lending further support  to the conclusions drawn from the core-level spectra [Fig.~\ref{ups}(a,b)]. 

 The stability of quasicrystals  has been related  to  the existence of a pseudogap in the electronic DOS  around the Fermi level  ($E_F$) originating from the  interaction between the Fermi surface and the quasi-Brillouin zone of the quasicrystal\cite{Nayak_prl12,Hafner92,Stadnik_prl96}. In the valence band spectrum of $i$-Al-Pd-Mn [Fig.~\ref{ups}(d)], the pseudogap was ascribed to  the rounded shape of the spectral function close to  $E_F$\cite{Stadnik_prl96}. With Sn deposition up to 1 ML (0.3 nm), we observe an interesting  change  in its shape [Fig.~\ref{ups}(e)]: the spectral weight decreases compared to  $i$-Al-Pd-Mn, indicating a deepening  of the pseudogap for the Sn monolayer, as  is evident from the green shaded region. The pseudogap is observed  up to the thickness of 2.3 nm (orange shaded region), as shown by the comparison  with a metal Fermi edge  in electrical contact with the specimen. However, for thicker layers, the pseudogap becomes shallower compared to the substrate, which is possibly related to the increase in disorder, as also shown by the  relative weakening of the LEED spots, as discussed earlier in Section III.C. % [Fig.~\ref{thickleed} and Fig.~S5]. 

The DOS for relaxed bulk and slab Sn clathrate is shown in Fig.~\ref{ups}(f). It is interesting to note that the bulk Sn quasiperiodic clathrate is semiconducting with a band gap of about 0.4~eV, in contrast to bulk $i$-Al-Pd-Mn that shows a pseudogap\cite{KH}. In contrast to bulk, the DOS for the slab of Sn quasiperiodic clathrate shows a pronounced pseudogap with the minimum at 0.3 eV (red arrow), the states near $E_F$ having almost equal contributions from the Sn $p$ and Sn $d$ states, while the contribution from the $s$ states is marginal. The pseudogap is deeper compared to $i$-Al-Pd-Mn, where the DOS was calculated for MS and M slabs\cite{KH}.   % There is much less - but still some - s-states proportion, too.}
~ Thus, the Sn clathrate model  explains the interesting observation from photoemission that the pseudogap is deeper for the Sn monolayer compared to the substrate $i$-Al-Pd-Mn.   
It is worth to note that the inter atomic bonding in  Sn  is  different from Pb, where a wider pseudogap was reported from scanning tunneling spectroscopy, but it  was later ascribed to the splitting of  Pb 6$p$ band due to  large spin-orbit coupling\cite{Ledieu_prb08,Krajci_PbAPM_prb10,Sharma13}. Pb does not have  $s-p$ hybridization since the $s$ and  $p$ bands are clearly separated, and thus the bonding is mediated by $p$ orbitals  only. Here, the deepening of the pseudogap around $E_F$ results from formation of the puckered clathrate structure with  high covalency and $sp^3$ bonding between the Sn atoms. The clathrate structures perfectly support the $sp^3$ bonding scheme, with exclusively tetrahedral local environments like in $\alpha$-Sn structure - but while in the latter the empty spaces are condensed into flat interlayer area, in clathrates they are isolated in four types of approximately spherical cages  (Fig.~\ref{cages}).

 \subsection	{The relaxed clathrate model and the STM motifs}

\begin{figure*}[tb] 
	\includegraphics[width=165mm,keepaspectratio]{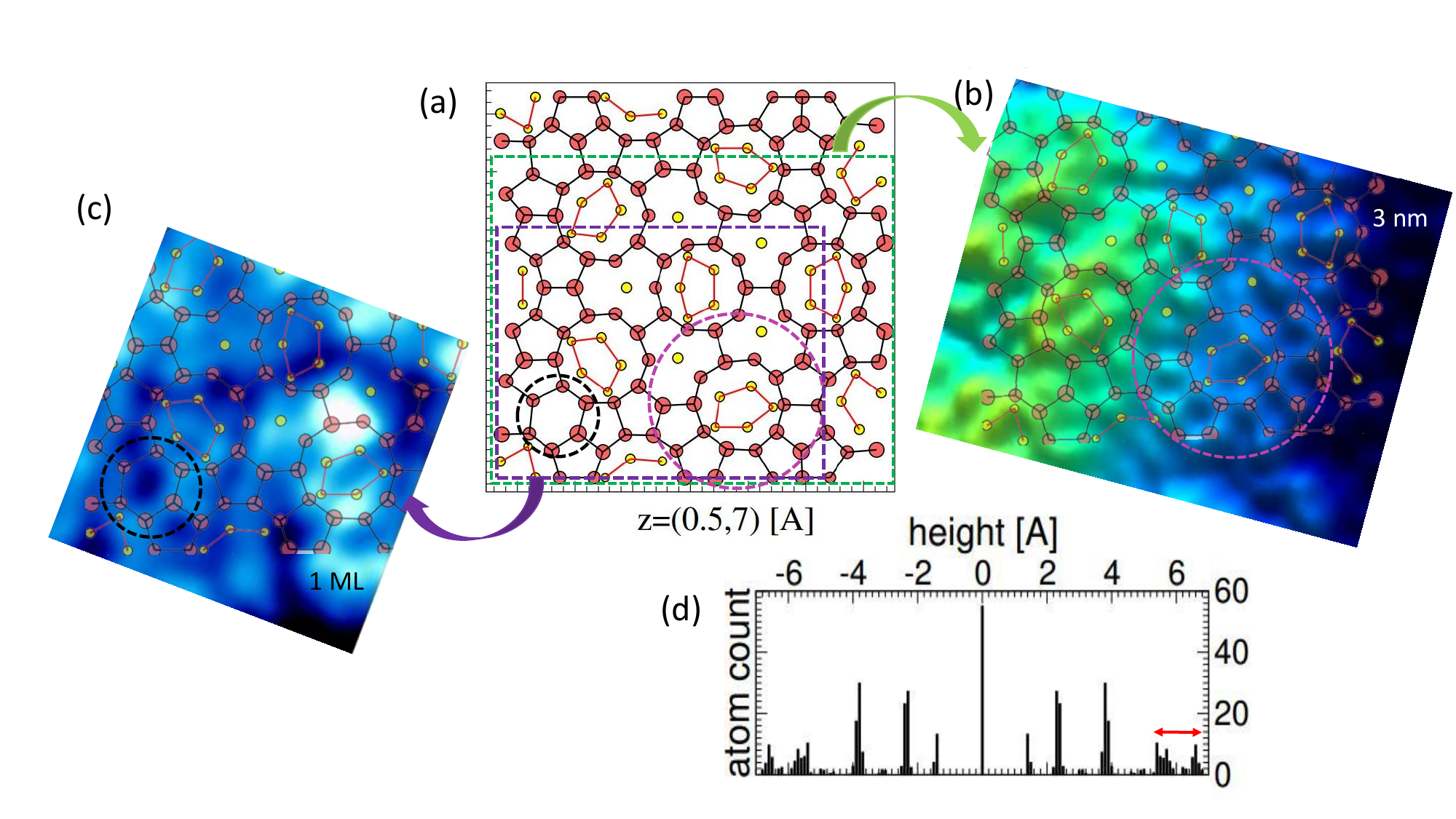} %{8_Sn_AlPdMn_2019_prr_final_figs_ver1_09}  %{Final_figures_Sn_AlPdMn_2019_ver8_fig3} 
	% Source:
	
	\vskip -5mm
	
	\caption{ (a) The most stable Sn clathrate slab fully optimized with surface relaxation using DFT for atom concentration of 0.35 atom/\AA$^2$ with dimensions of 35\AA$\times$\,35\AA. (b)  A part of the relaxed clathrate model  enclosed by a green dashed rectangle in (a) is overlaid on the STM image of 3 nm thick Sn layer   (size 16~nm$\times$\,13~nm, obtained from Fig.~\ref{dome}(d)  %(8nm x 8nm) 
			~that includes a crown motif highlighted by pink dashed circle)  %[see Fig.~\ref{dome}(d)] 
			 ~after $\tau$$^3$ scaling (see text). (c) A part  of the relaxed clathrate surface model  enclosed by violet dashed rectangle in (a) overlaid on the  STM image of  1 ML Sn (size 7~nm$\times$7~nm,  obtained from Fig.~\ref{stm}(a) that  %(4~nm$\times$\,3.5~nm) 
			~includes a hexagonal tile centered motif highlighted by a black dashed circle). %in Fig.~\ref{stm}(a) 
			 (d) Height histogram  of the Sn atoms within the slab, the red double arrow indicates the extent of puckering in the surface region.}
	\label{relaxclathrate}
\end{figure*}

%\section{The relaxed clathrate model and the STM motifs}
%\label{sec:stm-motifs}
The structure of the most optimal surface of the Sn clathrate obtained by relaxing the atom positions in our DFT calculation  is shown in Fig.~\ref{relaxclathrate}(a). We find that it exhibits protruding dodecagons projecting down on the tiling vertex positions, while the pentagonal caps of the dodecahedral cages suffer considerable distortion. The height distribution of the Sn atoms shows existence of different layers in the slab [Fig.~\ref{relaxclathrate}(d)]. However, their widths are considerable  and in particular, the surface region  shows that the spread in $z$ is $\approx$2\AA~ (red double arrow). This demonstrates that the surface is  corrugated, in agreement with  STM that shows large root mean square roughness (Section III.C). Such large corrugation arises because the clathrate structure fills the space very loosely and is made of large empty cages, and there may be wells even 4\AA~ deep. Moreover, in reality, all the atom positions of the layer are not occupied and disorder is present, in agreement with the results from LEED discussed in Section III.C. 

The relaxed clathrate structure is in good agreement with STM and the motifs are also reproduced.  For example, in Fig.~\ref{relaxclathrate}(b), a part of the clathrate structure enclosed by green dashed rectangle  is overlaid on a STM image  from 3 nm thick Sn layer that includes a crown motif (encircled by pink dashed circle)  after $\tau^3$ inflation and the agreement is found to be satisfying.  $\tau$-inflation is well known in quasicrystals due to their self-similar nature, for example, in $i$-Al-Pd-Mn, the fundamental intercluster linkage of 7.75\AA~ is $\tau$ inflated along the 2- fold direction, while $\tau$$^3$ inflation is observed along 5-fold or 3-fold directions.%\cite{a reference here} 
~$\tau$$^3$ inflation  has also been reported in a binary quasicrystal Yb-Cd with the formation of cluster of clusters\cite{Takakura07}.  Fig.~\ref{relaxclathrate}(c) provides another example, where a STM image containing a  hexagonal tile centered motif (encircled by black dashed circle) observed for 1 ML Sn is reproduced by a hexagon and the pentagons surrounding it from the violet dashed rectangle in Fig.~\ref{relaxclathrate}(a) with $\tau^2$ inflation.   Thus, the relaxed clathrate model is able to reproduce the STM images %of sizes as large as 16~nm$\times$13~nm 
	~[Fig.~\ref{relaxclathrate}(b, c)], which reconfirms the validity of this structural model.

\subsection{Nucleation and compatibility of the Sn clathrate structure with $i$-Al-Pd-Mn}

\begin{figure*}[tb] 
\includegraphics[width=175mm,keepaspectratio]{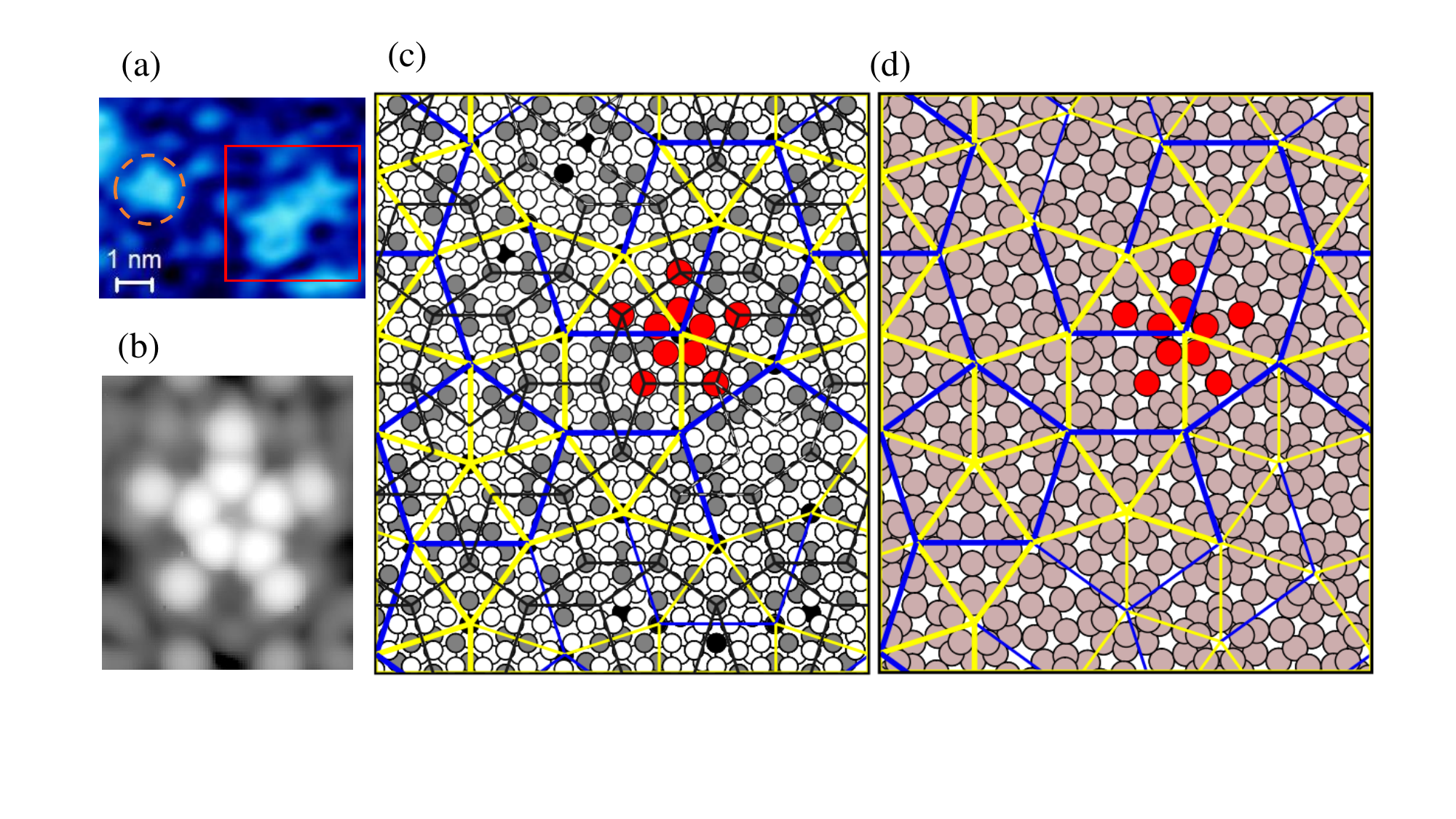}  %{Final_figures_Sn_AlPdMn_2019_ver8_fig3} 
	% Source:
	\vskip -10mm
	\caption{ (a) The STM image of  Sn white flower (SnWF) for 0.2 ML Sn on $i$-Al-Pd-Mn with $I_T$= 0.6~nA, $U_T$= 1.8~V. 
		%(b)  The calculated STM image of the SnWF cluster. The marked pentagon corresponds to the P1 tiling.
		~ (b) The calculated STM image of the
		SnWF cluster. The dimensions of the area are 20.3\AA\,$\times$\,23.9\AA. (c) The known atomic structure of the 5/3 approximant of $i$-Al-Pd-Mn surface
		(Al- open circles, Pd- gray circles, Mn- black circles) with dimensions of 53.2\AA\,$\times$\,62.5\AA.  Red circles show the atomic configuration of the SnWF
		(starfish) cluster. (d) The atomic structure of the quasiperiodic Sn clathrate corresponding to the 5/3 approximant. The configuration of the starfish cluster (red) is a part of the clathrate structure.} 
	
	\label{rtclathrate}
\end{figure*}

In this section, in order to find the genesis  of the  Sn clathrate structure and establish its compatibility with the substrate $i$-Al-Pd-Mn, we have studied the sub-monolayer coverages of Sn by STM to identify the nucleation centers. We find occurrence of pentagonal clusters that look like flowers with five bright petals at 0.2 ML coverage (encircled by orange dashes in [Fig.~\ref{rtclathrate}(a)], a larger area image is shown in Fig.~S8\cite{supplement}).    These clusters are formed on specific areas of the 5-fold surface recognized in STM images as white flowers\cite{KHLM}, and so we call them as the Sn white flower (SnWF) clusters. As shown in Fig.~S8 by orange circles, the SnWF's are  oriented in the same direction and are mostly isolated. However, we are able to identify regions, for example, inside the red rectangle in Fig.~\ref{rtclathrate}(a) that has two  SnWF's that are nearest neighbors.  

Similar five-fold clusters have also been observed for Pb\cite{LedieuPbAPMprb09} and Bi\cite{Smerdon_prb08}, which on the $i$-Al-Pd-Mn surface exhibit pseudomorphic growth up to a monolayer.  In both systems, these five-fold clusters have been dubbed as the {\it starfish} clusters.  On the $i$-Al-Pd-Mn surface, deposited Sn forms thick layer with the structure of quasicrystalline clathrate.  The SnWF clusters appearing at the sub-monolayer coverage play here an important role of nucleation centers for growth of the quasicrystalline layer.

To explain the role of the SnWF clusters we use the model of the 5/3 approximant of
$i$-Al-Pd-Mn\cite{KH,KWHKM}. The structure of the 5-fold surface that can be described by the
Penrose P1 tiling with an edge length of 7.76\AA\cite{KH} [black lines in
Fig.~\ref{rtclathrate}(c)]. The  tiling consists of pentagon, thin rhombus, star, and boat. The
bulk structure of $i$-Al-Pd-Mn can be interpreted in terms of the pseudo-Mackay (pMC)
clusters. %{\bf **MK* and the Bergman (BC) clusters *MK**} 
~On the 5-fold surface, the pentagonal tiles represent sections over these pMC clusters. The equatorial section of the pMC cluster has a Mn atom in the center of the pentagonal tile [Fig.~\ref{rtclathrate}(c)] and in the STM image this motif is seen as the white flower  mentioned above\cite{KHLM}. If the center of the pMC is deeper below the surface plane,  this part of the surface corresponds to another characteristic pattern known as the {\it dark star}\cite{KHLM}. In the STM image of the bare surface shown in Fig.~S1(c)\cite{supplement}, motifs of the white flower and the dark star can be well recognized.

At the submonolayer coverage, the SnWF cluster formed on the white flowers consists of 10 adatoms\cite{Krajci_PbAPM_prb10}, as shown by the red atoms in Fig.~\ref{rtclathrate}(c).  %**MK* The reason for the preferential occupation of the white flower sites by Sn atoms is the presence of the reactive Mn atoms. The interaction of Sn atoms with Mn is much stronger compared to Pd or Al atoms. The inner Sn pentagon of the SnWF cluster is formed around the central Mn atom.  Sn atoms of the outer pentagonal ring occupy vertices of the pentagonal P1 tile.  These Sn atoms are here bound to the Pd atoms inside small Al pentagons (edge of 2.96 \AA). These Pd atoms below the surface plane are the centers of the Bergman clusters \cite{Krajci_PbAPM_prb10}. *MK**} 
~The calculated STM image of the SnWF is shown in Fig.~\ref{rtclathrate}(b), where shape is similar to that  observed in the experiment [Fig.~\ref{rtclathrate}(a) and Fig.~S8].  The central pentagon around this would show up as bright  maxima in the STM images. 
To examine the compatibility of the SnWF with the clathrate structure, it is interesting to compare the configuration of 10 Sn adatoms forming the SnWF cluster [Fig.~\ref{rtclathrate}(c)](marked red) with the structure of the clathrate layer [Fig.~\ref{rtclathrate}(d)].  The clathrate structure consists of regular dodecahedral cages of Sn atoms.  The dodecahedra are centered at the vertices of the R-T tiling [Fig.~\ref{rtclathrate}(d)].  The top (or bottom) pentagonal face of the dodecahedron corresponds to the inner pentagon of the SnWF cluster.

The edges of the R-T tiling have two different lengths, the shorter  $y$ (yellow) and longer $b$ (blue).  In the $i$-Al-Pd-Mn structure, the length $b$= 12.55\AA~ corresponds to the distance between two pMC clusters along the two-fold directions.  In the clathrate layer, each of the two neighboring dodecahedral cages at the distance of the longer edge $b$ are connected over one Sn atom.  This atom corresponds to the Sn atom from the outer pentagonal ring of the SnWF cluster. The bonding topology of the SnWF cluster is thus closely related to the structure of the Sn-clathrate  and is a valid part of the clathrate structure.  The SnWF's  can thus serve as nucleation centers for growth of the quasiperiodic Sn adlayer with the clathrate structure.

On the other side, as the bulk structure of clathrate is different from the bulk structure of $i$-Al-Pd-Mn, there are also significant differences between the SnWF configurations on the $i$-Al-Pd-Mn surface and in the clathrate adlayer.  On the 5-fold Al-Pd-Mn surface, the SnWF cluster has essentially a planar structure and the interatomic bonding has a metallic character. All the SnWF clusters on the surface have only one pentagonal orientation.  In the atomic structure of the clathrate, the inner and outer pentagons of the SnWF clusters are at different heights.  The interatomic bonding has the covalent sp$^3$ character. They have both pentagonal orientations.  The compatibility of the interface between the $i$-Al-Pd-Mn surface and the clathrate adlayer helps to stabilize the clathrate structure and prevents its conversion to a more stable crystalline structure.

%{\bf **MK* The compatibility of the 5-fold $i$-Al-Pd-Mn surface with the structure of clathrate is not only in the bonding topology of the SnWF clusters but also in the scales and quasiperiodic ordering of both structures.  The dimensions of the atomic structure of the 5/3 approximant of the quasiperiodic Sn clathrate are, within the accuracy of 1\%, almost equal to the dimensions of the $i$-Al-Pd-Mn 5/3 approximant.  The quasiperiodic ordering of Sn atoms in the clathrate layer can be described by the R-T tiling [Fig.~\ref{rtclathrate}(d)]. It is interesting that a large part of the $i$-Al-Pd-Mn surface can also be described by the same R-T tiling. *MK**} 

Comparing Fig.~\ref{rtclathrate}(c) and (d), one can see that the arrangement of the tiles in a large part of the tilings on both models marked by thick blue and yellow lines are exactly the same. The parts of the R-T tilings on the $i$-Al-Pd-Mn and Sn clathrate surfaces with different arrangement of the tiles are marked by thinner lines. The reason why both tilings are not the same on the whole surface areas of the 5/3 approximants is the following: out of 4 tiles of the P1 tiling describing ordering of atoms on the $i$-Al-Pd-Mn surface, three of them - pentagon, thin rhombus  and star are compatible with the R-T tiling, but the boat tile contradicts the R-T ordering. On the 5/3 approximant model of the $i$-Al-Pd-Mn surface, there are 2 boat tiles [dashed black lines in Fig.~\ref{rtclathrate}(c)]. Around the boat tiles, the ordering of $i$-Al-Pd-Mn atoms can be again described by the R-T tiling (thin blue and yellow lines), however, with somewhat different arrangement of the tiles than in the clathrate model [Fig.~\ref{rtclathrate}(d)].  The boat tiles can thus induce defects in the R-T tiling.  The presence of the boat tiles can be the origin of the observed partial disorder in the Sn adlayer.

\section{CONCLUSIONS}

 Our work demonstrates formation of a new quasicrystalline structure in a thick Sn layer up to a thickness of atleast 4 nm.  Based on multiple evidences from both experiment and theory, we propose a quasiperiodic clathrate structural model for the Sn layer. 
   Sn  retains  quasicrystallinity up to a thickness that is highest reported so far for any element, and this is almost in the realm of bulk-like growth, where the influence of the substrate  potential is negligible. 
    The STM  motifs (crown, wheel and hexagonal tile centered motif in Fig.~\ref{stm}) and the LEED IV curves (Fig.~S3 of SM\cite{supplement}) of the Sn monolayer are  different from $i$-Al-Pd-Mn that is described by the Penrose P1 tiling. The motifs of the thick Sn layer are similar to the motifs of the first  layer. % showing that the ordering propagates from the first layer by mediation from layer to layer. 
   	~This shows a structural difference between the Sn layer  and the substrate, indicating that Sn grows with its intrinsic quasiperiodic structure and the substrate is unable to force a pseudomorphic growth. The role of the substrate is solely to suppress initial formation of any stable crystalline forms of Sn. 
   %%The motifs observed on the Sn layer by STM  are  clearly different  from %the dark star and white flower motifs of 
%%~the  well known motifs on the $i$-Al-Pd-Mn surface that are described by the Penrose P1 tiling. %For example, the wheel and crown motifs cannot be described by Penrose tiling.  Motifs with hexagonal tiles are also observed. Pentagons with bright vertices are different from the white flowers of $i$-Al-Pd-Mn because here a central dark region is observed. 
%%~The structural difference between the substrate and the Sn layer is also supported by LEED. 
~Photoemission spectroscopy shows that the pseudogap around the Fermi level  is deeper for the first Sn monolayer compared to  $i$-Al-Pd-Mn  and exists for the thick layer, indicating its stability.  
Using density functional theory,  a comparative study of the free slab energies shows that surface energy favors clathrate over $\alpha$-tin up to about 4 nm layer thickness, and matches $\beta$-tin for narrow window of slab thickness of 2-3 nm. The bulk  clathrate exhibits gap opening near Fermi energy, while the free slab form exhibits a pronouced pseudogap that explains the photoemission result. The STM images and the motifs observed on the Sn layer show good agreement with the proposed clathrate model.
%well reproduced  by the clathrate model. 
~The SnWF clusters constitute the nucleation centers for growth of the quasiperiodic Sn adlayer since these are  a valid part of the Sn clathrate structure. %{\bf On the basis of these evidences, we interpret the quasicrystalline layer of Sn as a clathrate quasicrystal.} 
~It is realized on the $i$-Al-Pd-Mn surface because the cluster-cluster separations (12.55~\AA) observed on the latter  %SRB: changed to 12.55 A from 1.28 nm)
~ enjoy magic consistency with the cage-cage separations in the Sn clathrate.  Last but not the least,  although here we do not observe Sn quasicrystal in the bulk form,  the lower bulk total energy of the quasiperiodic Sn clathrate  compared to  $\beta$-Sn suggests that %if $\beta$-Sn can exist above 286 K, 
~bulk quasicrystalline Sn  might also exist in some part of its phase diagram. %This leads us to speculate that tin is  the most suitable candidate in the search for elemental quasicrystallinity in the bulk form.

\section{ACKNOWLEDGMENTS}
		M. M. and 	M.K. thank support from the Slovak Grant Agency VEGA (No. 2/0082/17), and from APVV (No. 15-0621). Part of the calculations were performed in the Computing Center of the Slovak Academy of Sciences using the supercomputing infrastructure acquired 	under projects ITMS 26230120002 and 26210120002. K.P. thanks CSC–IT Center for Science and  the Academy of Finland under Grant No. 277829. Karsten Horn is thanked for his constructive suggestions and support. M. Maniraj is   grateful to the C.S.I.R, New Delhi for research fellowship. J. Nayak, P. Bhakuni and M. Balal are thanked for assistance during some of the experiments. 
		Work by TAL and DLS was supported by the U.S. Department of Energy (DOE), Office of Science, Basic Energy Sciences, Materials Science and Engineering Division. The $i$-Al-Pd-Mn substrate was synthesized at the Materials Preparation Center, at the Ames Laboratory, which is operated for the U.S. DOE by Iowa State University under contract  DE-AC02-07CH11358.\\

\noindent Electronic address of the corresponding authors:\\ $^{*}$Marek.Mihalkovic@savba.sk, \\
$^{**}$Marian.Krajci@savba.sk, \\$^{\ddagger}$barmansr@gmail.com

\setcounter{figure}{0}
\renewcommand{\figurename}{FIG.~S}

\newpage
%\begin{document}
	
	%\maketitle{\centerline {\em Supplementary material to the paper entitled}
	{{\begin{center} Supplementary material to the manuscript entitled:\\ 
				\bf\large  {Quasiperiodic ordering in thick Sn layer on $i$-Al-Pd-Mn:  A possible quasicrystalline clathrate}
				
	\end{center}}}

	~~\\
	~~\\
	\noindent {\bf The supplementary material contains  eight figures (S1$- $S8)  and  of four video files named “3nmleed" for  3 nm,   “1.8nmleed” for 1.8~nm, “1nmleed” for  1~nm  and  “0.25nmleed” for the 0.25~nm  or 1 ML Sn layer thicknesses. The video files show a sequence of LEED patterns  as a function of beam energy, as indicated in the top left corner for each pattern. The  video files are uploaded separately.}   
	%\newpage
	%\newpage
	\begin{figure*}[htb]
		%%	\centering
		%%	\epsfxsize170mm
		%%	\epsffile{Fig_S4_STM_Al-Pd-Mn_ver4.eps} 
		\includegraphics[width=170mm,keepaspectratio]{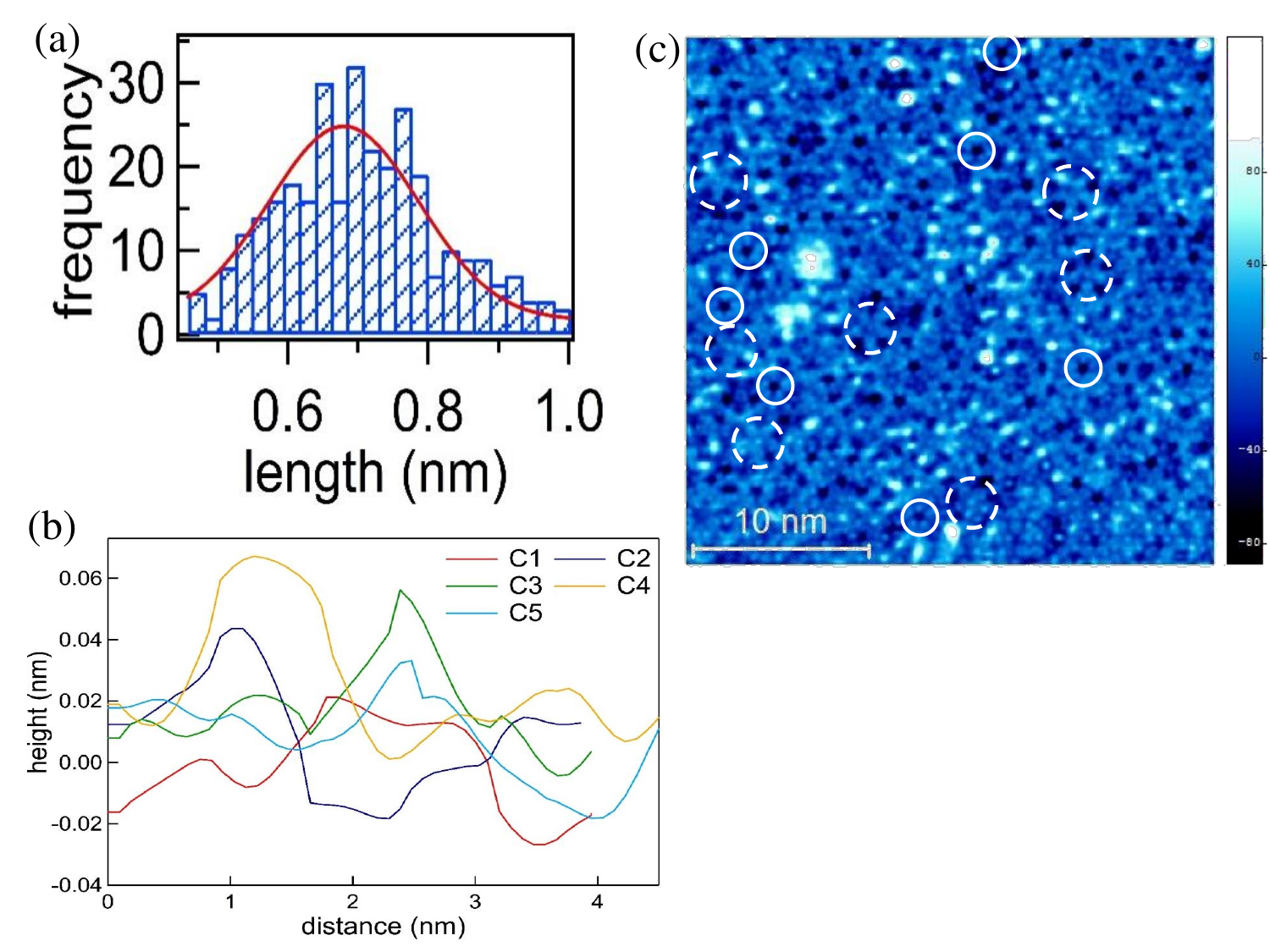}%{Fig_S4_STM_Al-Pd-Mn_ver6_pdf}
		\vskip -5mm
		\caption{(a) Distribution of the length of the sides of the pentagonal ($P$)-tile.  (b) Height  profiles along the sides of the $P$-tiles  in Figure~1 (e) of main manuscript.(c) High resolution STM image of substrate $i$-Al-Pd-Mn ($I_T$=\,0.1~nA, $U_T$=\,1.5~V). The white  dashed and white solid line circles enclose white flowers (WF) and dark stars (DS), respectively. 
		} 
		%{\bf D,} Constant current  high resolution STM  image with $I_T$= 0.2~nA tunneling current  and   $U_T$=~1.5~V bias voltage at RT  for  1st laver Sn after low-pass Fourier transform filtering. 
		%	~{\bf C,} High resolution STM image of substrate $i$-Al-Pd-Mn ($I_T$=\,0.1~nA, $U_T$=\,1.5~V). The white  dashed and white solid line circles enclose white flowers (WF) and dark stars (DS), respectively. 		{\bf D,}  The height distribution histogram of  Fig.1{\bf A} {\it i.e.} 1 ML Sn (green curve) and  bare $i$-Al-Pd-Mn in {\bf C} (red curve).   The larger corrugation of the Sn layer is  corroborated by  the  full width at half maximum (FWHM)  of the height distribution histogram recorded under similar tunneling conditions. FWHM  is 0.08~nm for the Sn layer, while  it is  smaller (0.05~nm) for  $i$-Al-Pd-Mn. } %  1 ML  Sn/$i$-Al-Pd-Mn   exhibits  larger corrugation that is  quantified by  the root mean square roughness ($S_q$): for the Sn layer  $S_q$ is 0.04~nm, whereas for $i$-Al-Pd-Mn it is 0.023~nm.}
		~~\\
		~~\\
		~~\\
		~~\\
		~~\\
		~~\\
		\label{S1}
	\end{figure*}
	~\\
	~\\
	\begin{figure*}[h!tb]
		%%	\centering
		%%	\epsfxsize=105mm
		%%	\epsffile{Fig_S1_ver3.eps}
		\includegraphics[width=190mm,keepaspectratio]{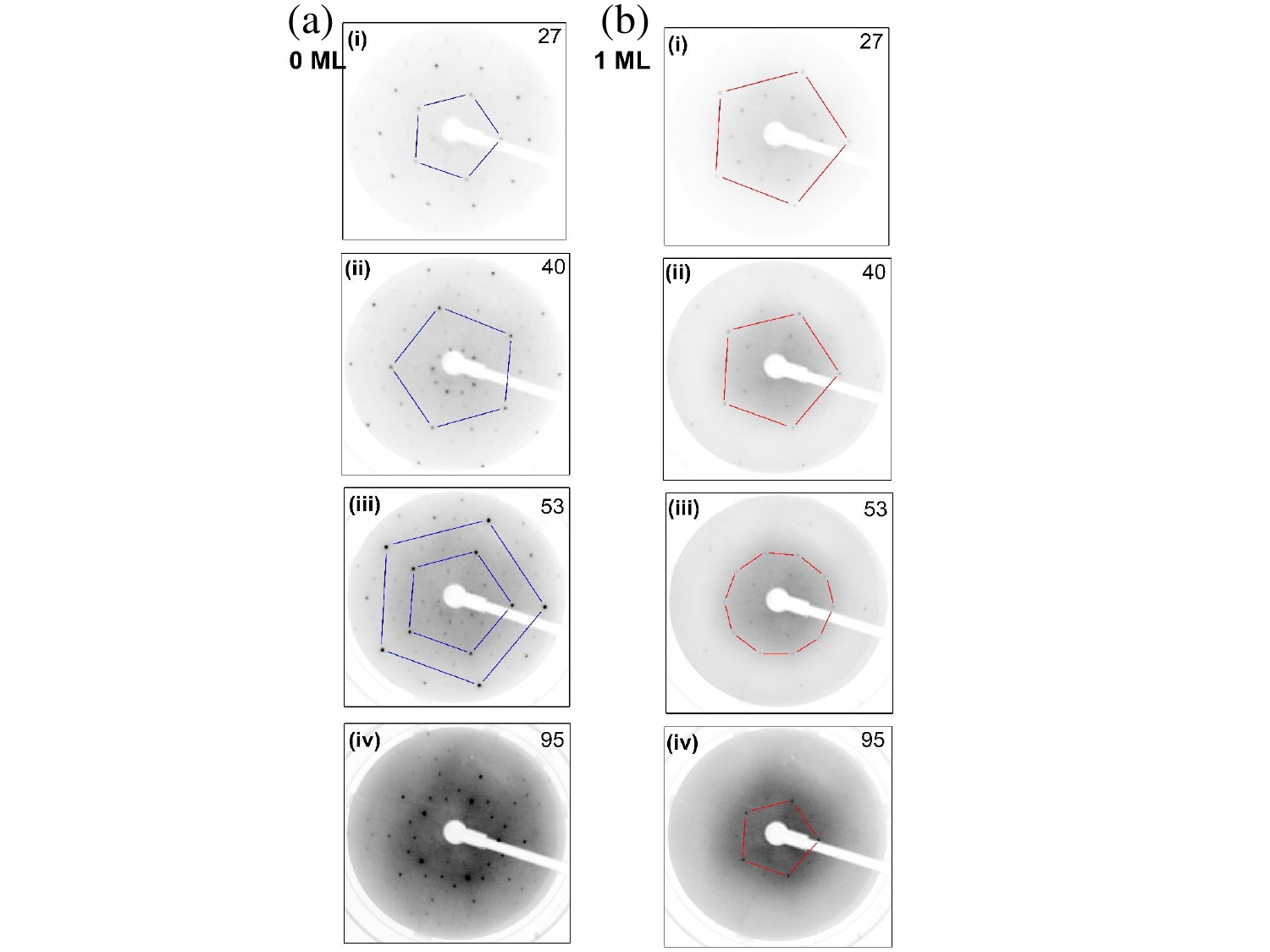} % Source: analysis/LEED_Auger/AlPdMn_Sn_LEED_PS.pxp, Layout2_3

		\caption{LEED patterns  of (a) $i$-Al-Pd-Mn and (b) 1 ML Sn/$i$-Al-Pd-Mn  for different beam energies (E$_p$), shown at the top right corners in eV. {\bf At $E_p$= 27 eV}, for $i$-Al-Pd-Mn, the inner set of spots form a  pentagon pointing downwards (blue lines), while the outer  ring has ten spots of almost similar intensity. For 1 ML Sn, the opposite is observed:   the inner ring has  ten almost similar  intensity spots, while the outer ring has five dominant spots forming a down pentagon (red lines) (a(i) and b(i)). {\bf At E$_p$= 40 eV}, ten spots of almost similar  intensities are observed 
			for $i$-Al-Pd-Mn in the innermost ring, while  1.1 ML Sn shows a down pentagon (red lines) that is larger in size.  In the outer set, an up pentagon (blue lines) in prominently observed in $i$-Al-Pd-Mn; whereas a down pentagon (red lines) is observed for Sn (a(ii) and b(ii)). 
			At {\bf E$_p$= 53 eV}, $i$-Al-Pd-Mn exhibits two dominant down pentagons (blue lines) (a(iii)), with two sets of 10 spots as inner rings.  Sn  exhibits an inner down pentagon and two outer pentagons  (red solid and dashed lines) (b(iii)). 
			At {\bf E$_p$= 95 eV} (a(iv) and b(iv)), only one prominent down pentagon (red lines) is observed.}
		
		\label{S2}
	\end{figure*}

	\begin{figure*}[h]
		%%	\centering
		%%	\epsfxsize=170mm
		%%	\epsffile{Fig_S2_IV_ver2.eps}
		\includegraphics[width=200mm,keepaspectratio]{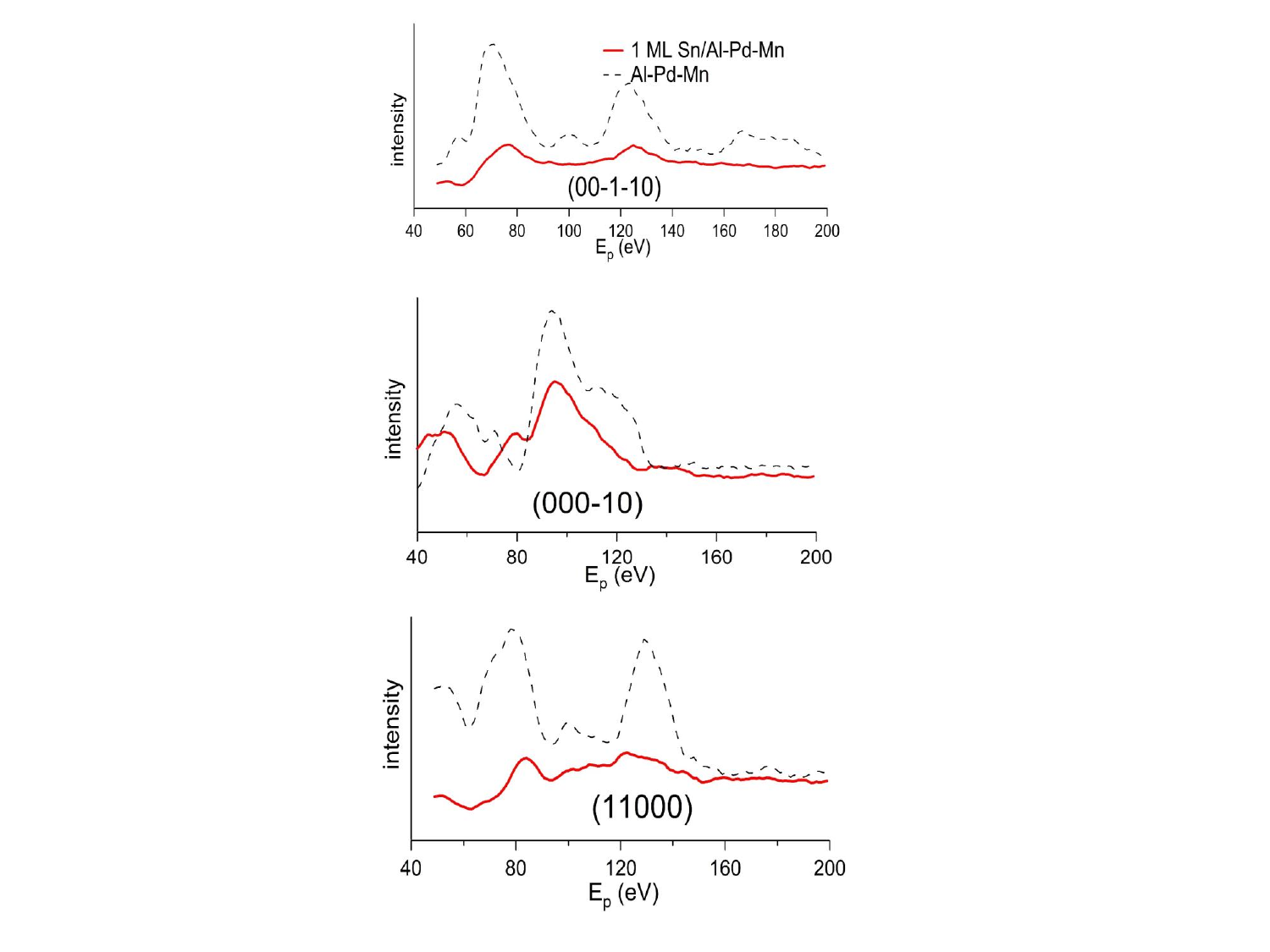} 
		\caption{Experimental IV curves for 1  ML Sn/$i$-Al-Pd-Mn compared with those of substrate $i$-Al-Pd-Mn for   (a) (00$\overline{1}$\,$\overline{1}$0), (b) (000$\overline{1}$0) and (c) (11000) LEED spots. 
			~~\\
			~~\\
			~~\\
			~~\\
			~~\\
			~~\\
			~~\\
			~~\\
			~~\\
			~~\\
			~~\\
			~~\\}
		\label{S3}
	\end{figure*}

	\begin{figure*}[htb]
		%%	\centering
		%%	\epsfxsize=105mm
		%%	\epsffile{Fig_S1_ver3.eps}
		\includegraphics[width=200mm,keepaspectratio]{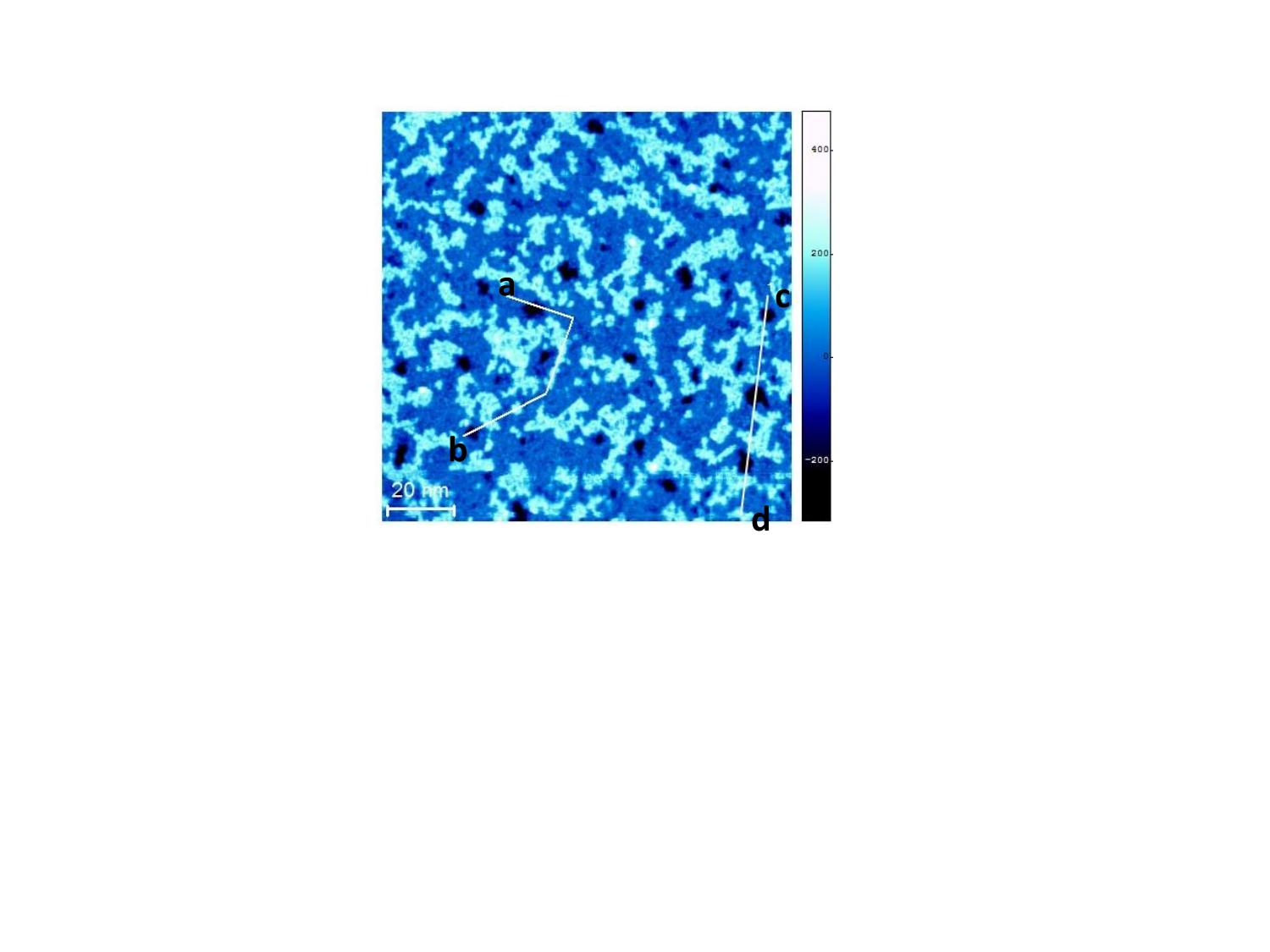} 
		% Source: analysis/STM/Figure/Sn_AlPdMN_4ML_fig_vks_ver6_srb.ppt; Slide 47 saved as pdf (using Save --> Pdf --> option --> choose slide) and the pdf file saved SA\supplementary\Sn_AlPdMn_4ML_fig_vks_ver6_srb_042.pdf

		%Source of A:  \analysis\STM\04-Oct-2014_10min Sn dep at 205C\[23-2]\[23-2]_PS\[23-2]_IB_350x350_bcrf_InvFFT_zoom_120x120.jpg  from SPIP loaded in STM/Figures/STM_PPT_02_12_2016_ver8_ps.ppt, Slide 28; inserted  circles and then loaded in Slide28, then  copied to Sn_AlPdMn_4ML_fig_vks_ver6_srb.ppt Slide26

		%Source of B:  \analysis\STM\Figures\STM_images for paper\20-Oct-2014_30min dep at 110C_SRB\[11-1]_PS\line profile\[11-1]ZTrUp_zoom_310x310_InvFFT_bcrf_zoom_258x203_polyS_wb from SPIP loaded in STM/Figures/STM_PPT_02_12_2016_ver7_ps.ppt, Slide 24; inserted yellow circles, line profile and then loaded in Slide 28,  copied to STM/Figures/Sn_AlPdMN_4ML_fig_vks_ver6_srb.ppt; Slide 27 for final figure.
		\vskip -4cm
		\caption{  An extended area STM topography image of 1.4 ML Sn/$i$-Al-Pd-Mn with $I_T$= 0.6~nA and bias voltage ($U_T$)= -2.6 V  at 300~K.} %after deposition at 430~K 
		%%%R1	{\bf B,} STM image with $I_T$= 0.5~nA, $U_T$)= -2.5 V showing pagoda-like motifs on the 2nd Sn layer.
		~\\ 
		~~\\
		~~\\
		~~\\
		~~\\
		~~\\
		~~\\
		~~\\
		~~\\
		~~\\
		~~\\
		~~\\
		~~\\
		~~\\
		~~\\
		~~\\
		~~\\
		
		\label{S2}
	\end{figure*}
	
	\begin{figure*}[htb]
		%%	\centering
		%%	\epsfxsize=105mm
		%%	\epsffile{Fig_S1_ver3.eps}
		\includegraphics[width=150mm,keepaspectratio]{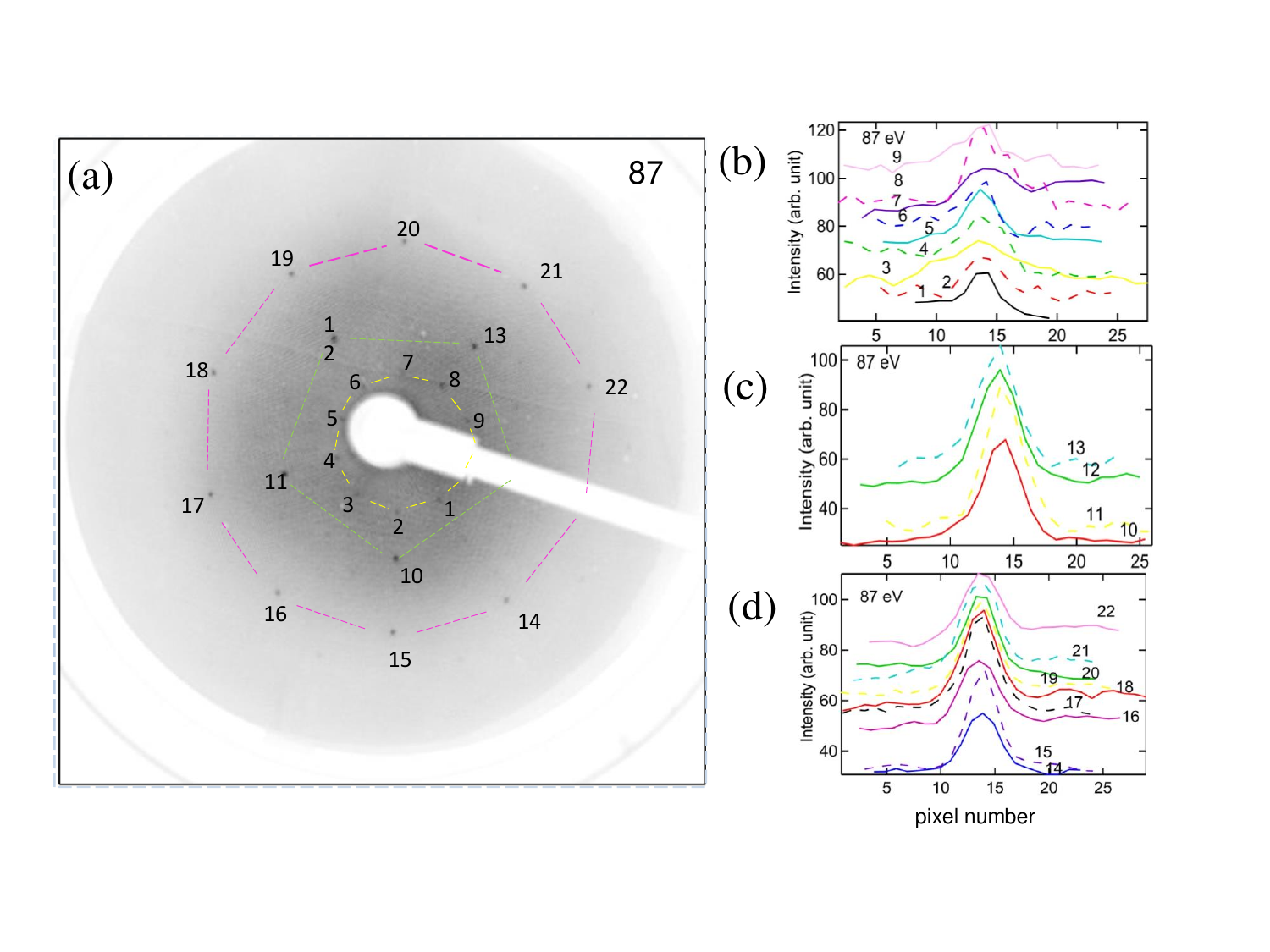} % 
		% Source: analysis/STM/Figure/Sn_AlPdMN_4ML_fig_vks_ver6_srb.ppt; Slide 23 saved as pdf (using Save --> Pdf --> option --> choose slide) and the pdf file saved in ...\Nature_Communications\4R_submission\supplementary\Sn_AlPdMn_4ML_fig_vks_ver6_srb_031_4thlayer_leed.pdf

		%Source of main figure same as Fig. 2H. 
		%Source of A-C: File supplied by VKS made from ...\analysis\2017\LEED\Sn_AlPdMn\2017_10_18_18102017_LT\5 min\5 min_LT_Rt annealing  from here bare 87 eV image loaded in imagej and inverted it and saved as tmp_87_5m_RT_anneal_imj.jpg now loaded as  ...\analysis\LEED_AUGER\Sn_AlPdMn_LEED_vks.pxp, graph 19 (main image) graph_28_prof_1 (A), graph_28_prof_2 (B), graph_28_prof_3 (C), converted them layouts 9-11, saved as jpg with names Layout9_87 eV_line profile_1.jpg, Layout10_87 eV_line profile_2.jpg, Layout11_87 eV_line profile_3.jpg, respectively;  then loaded in slide 17, Sn_AlPdMn_4ML_fig_vks_ver16.ppt;  copied to Sn_AlPdMn_4ML_fig_vks_ver6_srb.ppt Slide23 for final figure for 4R supplement. 	
		\caption{Intensity profiles through the (a) inner (b) intermediate and (c) outer set of spots for the LEED patterns of the 4th layer. The spots and the corresponding intensity profiles are numbered.}
		~\\ 
		~~\\
		~~\\
		~~\\
		~~\\
		~~\\
		~~\\
		~~\\
		~~\\
		~~\\
		~~\\
		~~\\
		\label{S3}
	\end{figure*} 
	\newpage
	\begin{figure*}[htb]
		%%	\centering
		%%	\epsfxsize=105mm
		%%	\epsffile{Fig_S1_ver3.eps}
		\includegraphics[width=125mm,keepaspectratio]{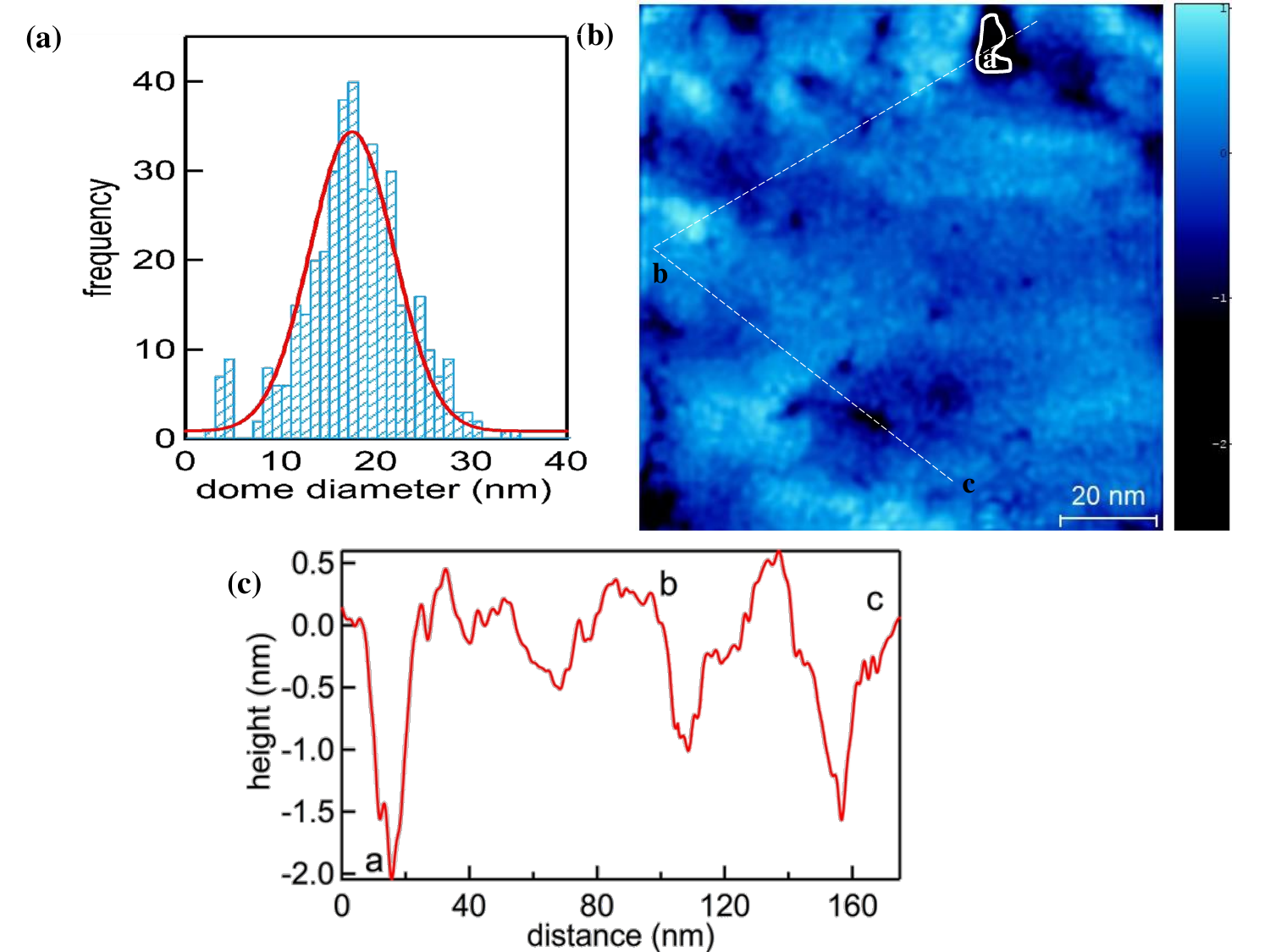} 
		% Source: analysis/STM/Figure/Sn_AlPdMN_4ML_fig_vks_ver6_srb.ppt; Slide 33 saved as pdf (using Save --> Pdf --> option --> choose slide) and the pdf file saved in ...\Nature_Communications\4R_submission\supplementary\Sn_AlPdMn_4ML_fig_vks_ver6_srb_33.pdf
		%	\vskip -4cm
		
		%Source of S6A: File supplied by VKS made from ....\08-Dec-2017_9.5min_LT dep_LT scan_tip_1(user tip)\analysis_srb_vks\[3-4]ZTrDn_vks\[3-4]  STM_Spectroscopy STM_zoom_500x448 Median RMS Filter 1x7_bcrf_InvFFT_3D.jpg, loaded in Slide 28, Sn_AlPdMn_4ML_fig_vks_ver16.ppt, copied to analysis/STM/Figure/Sn_AlPdMN_4ML_fig_vks_ver6_srb.ppt; Slide 33
		
		%Source of S6(B): File supplied by VKS made from ....\08-Dec-2017_9.5min_LT dep_LT scan_tip_1(user tip)\analysis_srb_vks\[20-9]ZReTrDn\[20-9]  STM_Spectroscopy STM_bcrf_InvFFT_bcrf_zoom_432x397.jpg, loaded in Slide 22 /Sn_AlPdMn_4ML_fig_vks_ver16.ppt, copied to analysis/STM/Figure/Sn_AlPdMN_4ML_fig_vks_ver6_srb.ppt; Slide 33
		
		%Source of S6(C): File supplied by VKS made from   ..\08-Dec-2017_9.5min_LT dep_LT scan_tip_1(user tip), \analysis_srb_vks\[20-9]ZReTrDn, saved from SPIP as [20-9]  STM_Spectroscopy STM_bcrf_InvFFT_bcrf_PolyLine Z(D).ASCII" and loaded in IGOR as  ...\analysis\STM\Figures\Sn_AlPdMn_hist_and_layout_vks_ver2.pxp, graph 18, converted in layout and saved, Layout6_[20-9]_poly line, loaded in Slide 22 of Sn_AlPdMn_4ML_fig_vks_ver16.ppt, copied to analysis/STM/Figure/Sn_AlPdMN_4ML_fig_vks_ver6_srb.ppt; Slide 33
		%\vskip -15mm
		\caption{ %An extended  area (120\,nm$\times$125\,nm) STM topography image corresponding to Fig. 3A of the manuscript for  3 nm thick Sn film at 80~K after deposition at 150~K showing the domes in 3D representation and the directions of the height profiles $abcd$ and $efgh$. {\bf B,}  
			(a) A distribution of the base diameter of the domes shown in Fig.~3(a) of the manuscript, the red line represents a Gaussian fitted to this distribution. (b) An extended area STM topography image corresponding to Fig.~3(d)  showing the directions of the height profile $abc$.  (c) Height profile along $abc$, where  the minimum region around $a$ with height $\leq$2nm is enclosed by a white curved line.} %{\bf C,} Fourier transform of  {\bf A} showing 4 weak spots that are encircled. These spots lie on the same circle and the angle between them that is close to the expected value of 36$^{\circ}$. The deviation is because of possible thermal drift in the STM image, as mentioned in the manuscript.  	
		~~\\
		~~\\
		~~\\
		~~\\
		~~\\
		~~\\
		~~\\
		~~\\
		~~\\
		~~\\
		\label{S6}
	\end{figure*}

	\begin{figure*}[htb]
		%%	\centering
		%%	\epsfxsize=105mm
		%%	\epsffile{Fig_S1_ver3.eps}
		\includegraphics[width=175mm,keepaspectratio]{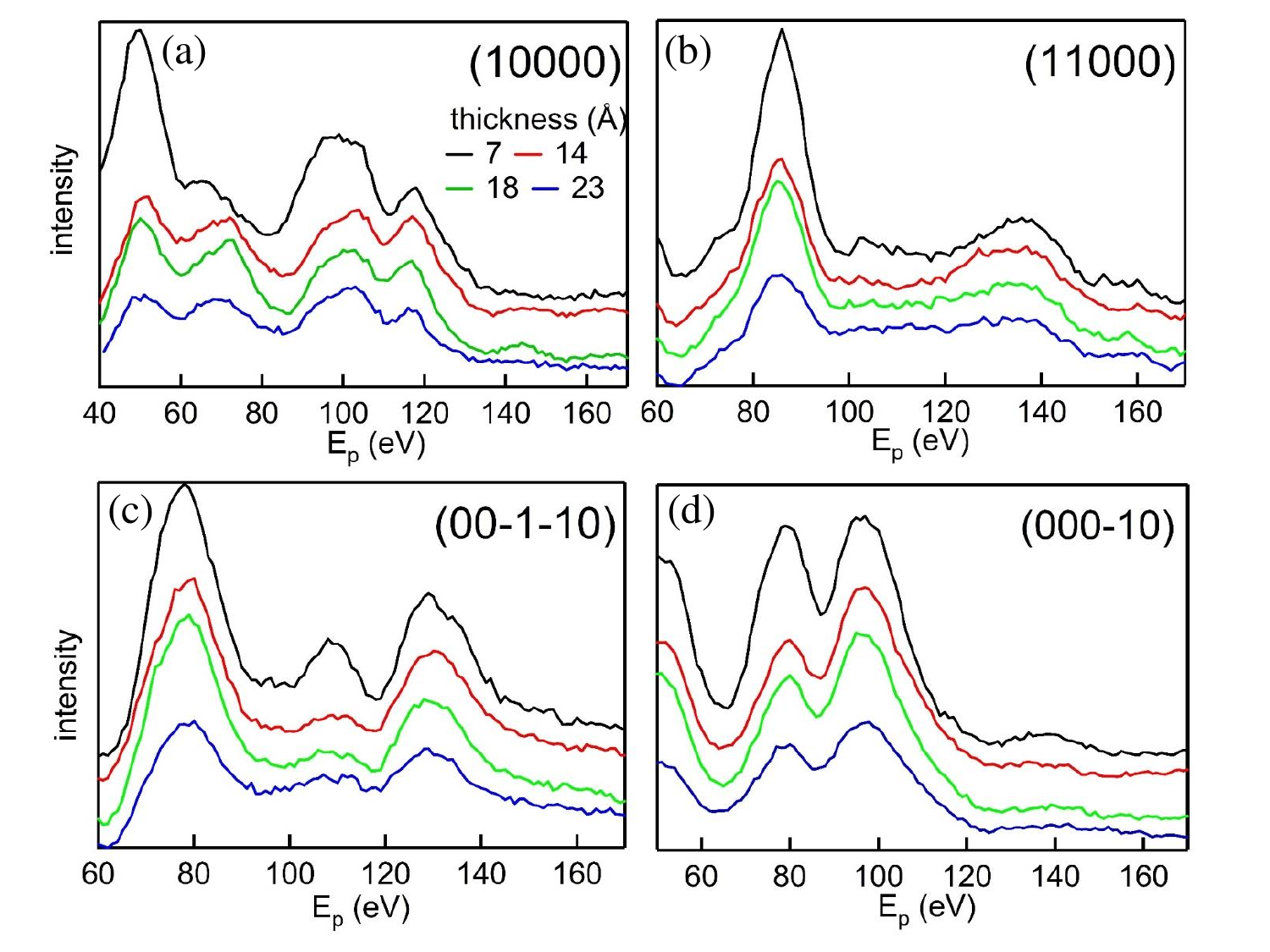} 
		% Source: analysis/STM/Figure/Sn_AlPdMN_4ML_fig_vks_ver6_srb.ppt; Slide 22 saved as pdf  and the pdf file saved in ...\Nature_Communications\4R_submission\supplementary\Sn_AlPdMn_4ML_fig_vks_ver6_srb_030_3nm_leed.pdf
		
		%Source of S7 main figures: File supplied by VKS made from ....\analysis\2017\LEED\Sn_AlPdMn\2017_12_08_08122017_9.5min at LT\analysis_vks  from here bare 46eV, 87 eV, 135 eV images loaded in Imagej and did twice background subtraction(light background and sliding paraboloid)+ 7 times auto contrast (for 46 eV) and 6 times auto contrast(for 87 eV and 135 eV) and inverted it then images loaded in photoshop, deleted dirt specs, and saved as tmp_46_imj_ps_9.5m_LT.jpg, tmp_87_imj_v2_ps.jpg, tmp_135_imj_ps_9.5_LT.jpg;  now these images loaded in \analysis\STM\Figures\Sn_AlPdMn_hist_and_layout_vks_ver2.pxp, graph 20, 21, 23, 24_Prof_1, 26_Prof_1, 27_Prof_1, 27_Prof_2 converted them layout (Layout2_87 eV, Layout6_46 eV_line profile, Layout7_87 eV_line profile_1, Layout8_87 eV_line profile_2, tmp_46_imj_ps_9.5_LT, tmp_135_imj_ps_9.5_LT, Layout5_135 eV_line profile) and saved them as jpg in \LEED\Sn_AlPdMn\08122017_9.5min at LT\vks_analysis, then loaded in Sn_AlPdMn_4ML_fig_vks_ver16.ppt; slide 19, copied to analysis/STM/Figure/Sn_AlPdMN_4ML_fig_vks_ver6_srb.ppt; Slide 22.
		
		%SOURCE : analysis/STM/Figures/Sn_AlPdMN_4ML_fig_vks_ver6_srb.ppt; Slide 49 saved 	\vskip -1cm
		\caption{Experimental IV curves for  Sn adlayers on $i$-Al-Pd-Mn as a function of thickness  for  (10000), (11000),  (00$\overline{1}$\,$\overline{1}$0) and (000$\overline{1}$0) LEED spots. The curves are staggered along the vertical axis. }
		~~\\
		~~\\
		~~\\
		~~\\
		~~\\
		~~\\
		~~\\
		~~\\
		~~\\
		~~\\
		
		\label{S7}
	\end{figure*}

	\newpage
	
	\begin{figure*}[htb]
		%%	\centering
		%%	\epsfxsize=105mm
		%%	\epsffile{Fig_S1_ver3.eps}
		\includegraphics[width=175mm,keepaspectratio]{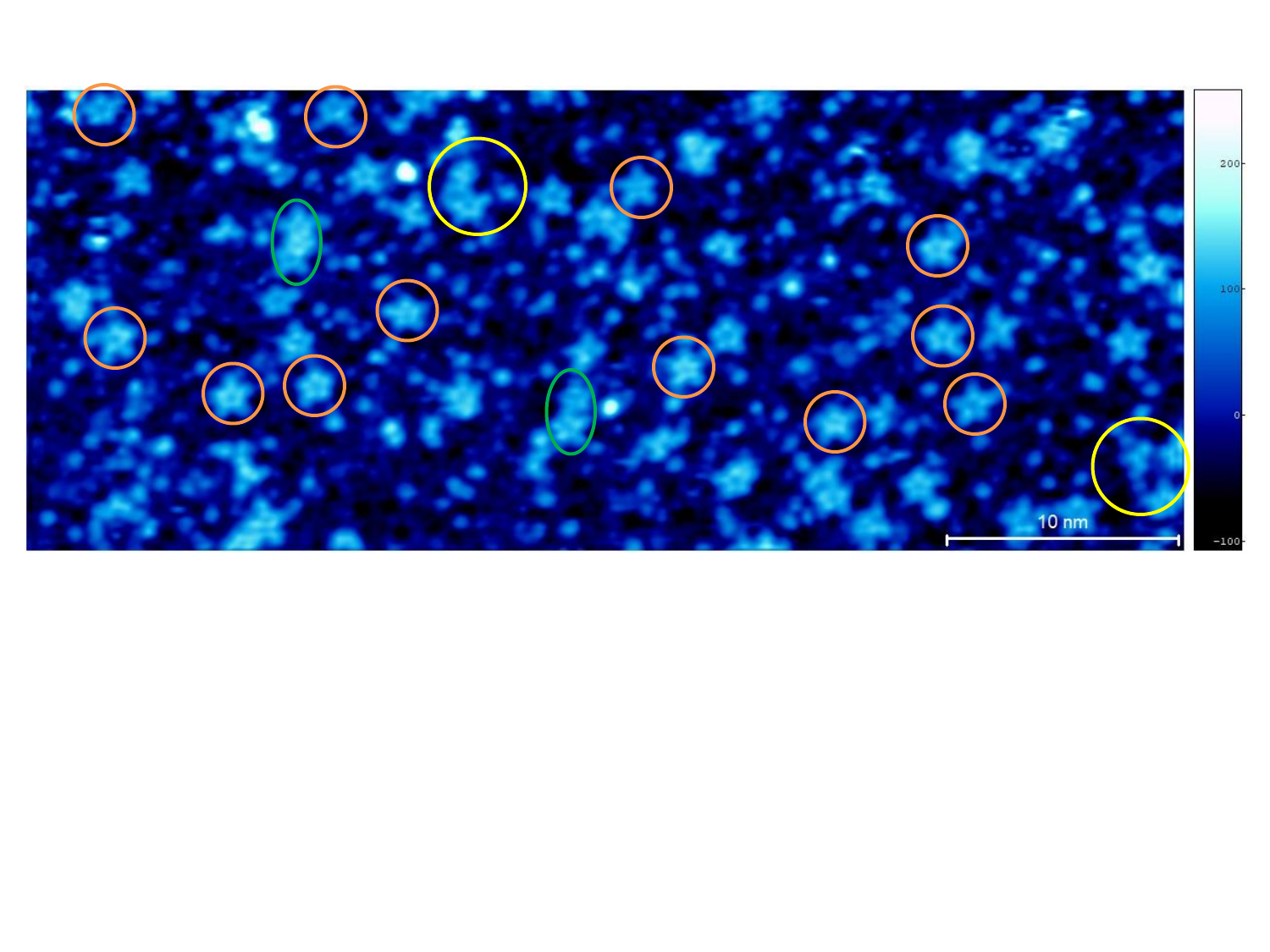} 
		\vskip -5cm
		\caption{An extended area STM image of  Sn white flower (SnWF) for 0.2 ML Sn/$i$-Al-Pd-Mn with $I_T$= 0.5~nA, $U_T$= 1.5~V showing n.n. SnWf's (green oval). Isolated single SnWF's are enclosed by orange circles and the multi SnWF and $P$-tile complex structures are highlighted by yellow circles. }
		\label{S8}
	\end{figure*}
	\newpage

\end{document}